\documentclass[prd,aps,
nofootinbib,floatfix,
superscriptaddress]{revtex4}
\usepackage{graphicx}
\usepackage{epsfig}
\usepackage{rotating}
\usepackage{amssymb}
\usepackage{subfigure}
\usepackage{dsfont}
\usepackage{psfrag}
\usepackage{amsmath,euscript,array,mathrsfs}
\usepackage{slashed}
\usepackage{array}
\usepackage{youngtab}
\usepackage{color}
\usepackage{bbold}

\usepackage{hyperref}
\hypersetup{
colorlinks = true,
linkcolor = blue,
linktocpage = true,
citecolor = blue
}

\allowdisplaybreaks

\newcommand{\beq}{\begin{equation}}
\newcommand{\eeq}{\end{equation}}
\newcommand{\beqs}{\begin{eqnarray}}
\newcommand{\eeqs}{\end{eqnarray}}

\renewcommand{\r}{{\rho}}

\newcommand{\dd}{\mbox{d}}

\newcommand{\be}{\begin{equation}}
\newcommand{\ee}{\end{equation}}
\newcommand{\ba}{\begin{array}}
\newcommand{\ea}{\end{array}}

\newcommand{\BigCurlyLeft}{\left\{ \rule{-0.1cm}{0.75cm} \right.}
\newcommand{\BigCurlyRight}{\left. \rule{-0.1cm}{0.75cm} \right\}}

\begin{document}

\title{Phase transitions and
light scalars
in bottom-up holography}
\author{Daniel Elander}
\affiliation{Laboratoire Charles Coulomb (L2C), University of Montpellier, CNRS, Montpellier, France}
\author{Ali Fatemiabhari}
\email{a.fatemiabhari.2127756@swansea.ac.uk}
\affiliation{Department of Physics, Faculty of Science and Engineering,
Swansea University,
Singleton Park, SA2 8PP, Swansea, Wales, UK}
\author{Maurizio Piai}
\affiliation{Department of Physics, Faculty of Science and Engineering,
Swansea University,
Singleton Park, SA2 8PP, Swansea, Wales, UK}

\date{\today}

\begin{abstract}

Within the bottom-up approach to holography, we construct a class of six-dimensional gravity models, and discuss solutions that can be interpreted, asymptotically in the far UV, in terms of dual five-dimensional conformal field theories deformed by a single scalar operator. We treat the scaling dimension of such operator, related to the mass of the one scalar field in the gravity theory, as a free parameter. One dimension in the regular geometry is compactified on a shrinking circle, hence mimicking confinement in the resulting dual four-dimensional theories.

We study the mass spectrum of bosonic states. The lightest state in this spectrum is a scalar particle. Along the regular (confining) branch of solutions, we find the presence of a tachyonic instability in part of the parameter space, reached by a smooth deformation of the mass spectrum, as a function of the boundary value of the background scalar field in the gravity theory. In a region of parameter space nearby the tachyonic one, the lightest scalar particle can be interpreted as an approximate dilaton, sourced by the trace of the stress-energy tensor, and its mass is parametrically suppressed.

We also compute the free energy, along several branches of gravity solutions. We find that both the dilatonic and tachyonic regions of parameter space, identified along the branch of confining solutions, are hidden behind a first-order phase transition, so that they are not realised as stable solutions, irrespectively of the scaling dimension of the deforming field-theory operator. The (approximate) dilaton, in particular, appears in metastable solutions. Yet, the mass of the lightest state, computed close to the phase transition, is (mildly) suppressed. This feature is amplified when the (free) parameter controlling the scaling dimension of the deformation is $5/2$, half the dimension of space-time in the field theory.

\end{abstract}

\maketitle

\tableofcontents

\section{Introduction}
\label{Sec:introduction}

Investigations into the  fundamental nature and origin of the Higgs boson, the latest
particle of the Standard Model (SM) to have been discovered~\cite{Aad:2012tfa,Chatrchyan:2012xdj},
are among the most topical in theoretical physics, in view of the ongoing precision experimental programme at
the Large Hadron Collider. 
Because of the approximate (explicitly and spontaneously broken) scale invariant nature of the SM theory,
the Higgs boson itself is in fact a dilaton, the pseudo-Nambu-Goldstone boson
associated with dilatations. It is natural to question whether this is just an accidental symmetry in the SM, 
or whether it has a more fundamental, important origin and role~\cite{Goldberger:2008zz}. In particular, is it possible for 
the Higgs boson  to originate as a composite dilaton in a more fundamental theory?
And if so, what type of underlying dynamics would yield  realistic values for its mass and couplings?

The literature on the dilaton effective field theory dates back many decades~\cite{Migdal:1982jp,Coleman:1985rnk},
and its applications have been long discussed in the context of
dynamical electroweak symmetry
 breaking~\cite{Leung:1985sn,Bardeen:1985sm,Yamawaki:1985zg}, extensions of the SM~\cite{Hong:2004td,Dietrich:2005jn,Hashimoto:2010nw,
Appelquist:2010gy,Vecchi:2010gj,Chacko:2012sy,
 Bellazzini:2012vz,Bellazzini:2013fga,Abe:2012eu,Eichten:2012qb,Hernandez-Leon:2017kea},
and, more recently, the interpretation of lattice data~\cite{Matsuzaki:2013eva,
Golterman:2016lsd,Kasai:2016ifi,Hansen:2016fri,Golterman:2016cdd,Appelquist:2017wcg,Appelquist:2017vyy,
Golterman:2018mfm,Cata:2019edh,Appelquist:2019lgk,Cata:2018wzl,Golterman:2020tdq,
Golterman:2020utm,LatticeStrongDynamicsLSD:2021gmp,Appelquist:2020bqj,Appelquist:2022qgl,Appelquist:2022mjb},
in particular in view of the numerical  work on certain $SU(3)$ gauge theories that
show indications of a light scalar bound state in the spectrum~\cite{Aoki:2014oha,Appelquist:2016viq, 
 Aoki:2016wnc,Gasbarro:2017fmi,Appelquist:2018yqe,Hasenfratz:2022qan,Fodor:2012ty,Fodor:2015vwa,
 Fodor:2016pls,Fodor:2017nlp,Fodor:2019vmw,Fodor:2020niv}.
 
The context of
gauge-gravity dualities~\cite{Maldacena:1997re,Gubser:1998bc,Witten:1998qj,Aharony:1999ti}
is particularly suitable for describing the dilaton and its dynamical origin,
both in its
bottom-up~\cite{Goldberger:1999uk,DeWolfe:1999cp,Goldberger:1999un,Csaki:2000zn,
ArkaniHamed:2000ds,Rattazzi:2000hs,Kofman:2004tk,Elander:2011aa,Kutasov:2012uq,Lawrance:2012cg,
Elander:2012fk,Goykhman:2012az,Evans:2013vca,Megias:2014iwa,Elander:2015asa}
as well as  top-down~\cite{Elander:2009pk,Elander:2012yh,Elander:2014ola,Elander:2017cle,Elander:2017hyr}
realisations.
In what follows, we restrict our attention to holographic duals
in which the geometry of a completely smooth classical background in
 a higher-dimensional gravity theory contains a shrinking circle,
the shrinking of which can be used to mimic
confinement in the dual field theory, as suggested in Ref.~\cite{Witten:1998zw}---see also Refs.~\cite{Sakai:2004cn,Sakai:2005yt,Brower:2000rp,Wen:2004qh,
 Kuperstein:2004yf,Elander:2013jqa,Elander:2018aub}.
The spectra of bound states of the strongly-coupled theory
can be computed perturbatively in its gravity dual, by exploiting the holographic 
gauge-invariant formalism developed in  Refs.~\cite{Bianchi:2003ug,
Berg:2005pd,Berg:2006xy,Elander:2009bm,Elander:2010wd}---see
also Refs.~\cite{Elander:2014ola,Elander:2017cle,Elander:2017hyr,Elander:2018aub,Elander:2020csd}.
 A useful diagnostic tool to identify (approximate) dilatons within such spectra
is provided by the probe approximation, as discussed in Ref.~\cite{Elander:2020csd}.
Furthermore, one can apply 
holographic renormalisation~\cite{Bianchi:2001kw,
Skenderis:2002wp,Papadimitriou:2004ap} to compute 
field-theory {quantities}, such as the free energy. 
We also find it appealing to adopt a simple scale-setting procedure, 
such as that proposed in Ref.~\cite{Csaki:2000cx}.

Among all possible realisations of the dilaton scenario,
we concentrate on the ideas developed in  
 Refs.~\cite{Kaplan:2009kr,Gorbenko:2018ncu,
Gorbenko:2018dtm,Pomarol:2019aae}---see also Ref.~\cite{CruzRojas:2023jhw} and references therein. 
Namely, we want to understand whether
a light dilaton state might arise in strongly-coupled theories,
the renormalisation group (RG) flow of which brings them in close proximity of 
a tachyonic instability. 
In the holographic dual 
description of conformal field theories (CFTs)---fixed points of the RG flow---the notion of the 
Breitenlohner-Freedman (BF)  bound~\cite{Breitenlohner:1982jf} captures 
the type of instability of interest.
In the case of confining theories, one can generalise this notion,
and ultimately arrive to the following, similar conclusions for physical observables~\cite{Elander:2020ial,Elander:2020fmv,Elander:2021wkc}.
In summary, the classical instability  is resolved by the presence of a first-order phase transition.
The unstable region of parameter space cannot be physically realised, 
which protects the physical spectrum from developing a tachyon.
Nevertheless, {one may ask whether,} for choices of parameters close to the phase transition, the lightest (neutral) scalar in the spectrum has {mass and
properties} that are influenced by the presence of the tachyonic instability itself,
even if the latter is only present in an unphysical region of parameter space. 

The analysis of  Refs.~\cite{Elander:2020ial,Elander:2020fmv,Elander:2021wkc}
indicates that it is possible to realise this scenario within well-established 
examples of  top-down holographic models, derived 
as classical solutions of  supergravities in various dimensions. 
In all three cases analysed so far, it has been found that for choices of the parameters that are close
to the phase transition boundary, the lightest scalar state is not parameterically light, as was predicted in Ref.~\cite{Pomarol:2019aae},
on the basis of the results from a bottom-up model obtained in a dedicated construction---for models in which no light dilaton is present, see for instance Refs.~\cite{Arean:2012mq,Arean:2013tja}. 

In this paper, inspired by Refs.~\cite{Elander:2020ial,Elander:2020fmv,Elander:2021wkc},
we build a class of bottom-up models that combine
the quadratic superpotential adopted in Ref.~\cite{DeWolfe:1999cp},
and the confinement mechanism in Ref.~\cite{Witten:1998zw}.
We assume that the holographic description of the dynamics
be encapsulated in a {model consisting of one real scalar}, coupled to gravity in $D=6$ dimensions, one of which is compacified on a circle.
Along the interesting, physical branch of solutions the circle shrinks to a point
at a finite position along the holographic dimension, intruducing a dynamical scale,
but the geometry is regular and smooth everywhere. 
What results is the dual description of a putative family of four-dimensional, confining theories,
that at short distances are best described by the circle compactification
of a five-dimensional CFT
in the presence of an operator with non-trivial dimension given by ${\rm max}(\Delta,5-\Delta)$, where $\Delta$ is a free parameter appearing in the scalar potential of the gravity theory.

The generic behaviour of the model, for each value of $\Delta$, is {qualitatively similar} to that found in Refs.~\cite{Elander:2020ial,Elander:2020fmv,Elander:2021wkc}, and described above, as we shall show.  The advantage of the bottom-up approach---aside from having a much simpler bosonic action---is that it allows us to vary $\Delta$. We are interested in the relation between $\Delta$ and the mass spectrum, in particular for the lightest scalar, and in proximity of the {phase} transition.

The paper is organised as follows.
We start by defining the gravity theory in $D=6$ dimensions, in Sec.~\ref{Sec:model}.
We present several classes of classical solutions in Sec.~\ref{Sec:solutions}, and then study the {spectrum of small fluctuations for one of these classes, the regular (confining) branch of solutions, in Sec.~\ref{Sec:glueballs}. The free energy of the different classes of solutions is discussed in Sec.~\ref{Sec:freeenergy}, where we analyse which branch of solutions is energetically favoured, depending on the parameters of the model.} We summarise the most salient numerical results in  Sec.~\ref{Sec:summary}, and outline future lines of research in Sec.~\ref{Sec:outlook}, while relegating to the Appendices many useful technical details.

\section{The model}
\label{Sec:model}

This section introduces a model, built within the context of bottom-up holography, that describes 
a real scalar field ${\phi}$ coupled to gravity in $D=6$ dimensions, with a simple quartic scalar potential~\cite{DeWolfe:1999cp}. The scalar field captures, in the gravity language,  the effects of the deformation of the dual five-dimensional CFT by a scalar operator, as well as the formation of the corresponding condensate. Furthermore, one of the  dimensions is compactified on a circle (in both field theory as well as gravity). As we shall see in Sec.~\ref{Sec:solutions}, this 
system admits solutions in which the circle shrinks smoothly at the end of space in the IR, thus introducing a physical low-energy scale, while mimicking (in analogy with top-down models) the effect of confinement in what would be the dual, four-dimensional field theory~\cite{Witten:1998zw}. We also find it convenient to provide the description
of  the gravity theory in $D=5$ dimensions, obtained after reduction on the circle.
  
\subsection{Action in six dimensions}
\label{Sec:6}
 
We follow the conventions in Ref.~\cite{Elander:2020csd} (see also references therein),
which we summarise in Appendix~\ref{Sec:DAction}.
 In order for the model to mimic the dual of a four-dimensional confining theory, we choose to work in 
$D=6$ dimensions.
The action is the following 
\begin{align}
	\mathcal S_6 &= \mathcal S_6^{(bulk)} + \sum_{i=1,2} \mathcal S_{5,i} \,, \\
	\mathcal S_6^{(bulk)} &= \int \dd^6 x \sqrt{-\hat g_6} \, \bigg\{ \frac{\mathcal R_6}{4} -
	 \frac{1}{2} \hat g^{\hat M \hat N} \partial_{\hat M} {\phi}  \partial_{\hat N} \phi 
	 - {\mathcal V_6(\phi)}  \bigg\} \,, \\
	\mathcal S_{5,i} &= (-)^i \int \dd^5 x \sqrt{-\tilde{\hat g}} \, \bigg\{ \frac{\mathcal K}{2} + \lambda_i(\phi) + f_i\left( \tilde{\hat g}_{\hat M \hat N} \right) \bigg\} \bigg|_{\rho = \rho_i} \,,\label{Eq:bbcc}
\end{align}
and besides the bulk part, $\mathcal S_6^{(bulk)}$, it contains also two boundary actions, $\mathcal S_{5,i}$, localised at the boundaries of the radial coordinate $ \rho_{1}<\rho < \rho_{2}$. The space-time index is $\hat M = 0,1,2,3,5,6$. 
The extrinsic curvature $\mathcal K$, 
 appearing in the Gibbons-Hawking-York (GHY) term of the boundary actions,
 depends on the induced metric on the boundaries, denoted $\tilde{\hat g}_{\hat M \hat N}$.

For the bulk potential $\mathcal V_6$, we choose to write 
\beqs
\label{eq:VfromW}
	\mathcal V_6(\phi) &= &\frac{1}{2} \left(\frac{\partial {\mathcal W}_6(\phi)}
	{\partial{\phi}}\right)^2	
	- \frac{5}{4} {\mathcal W}_6(\phi) ^2 \nonumber \\
 &=& -5 - \frac{\Delta (5 - \Delta)}{2} {\phi}^2 - \frac{5 \Delta^2}{16} {\phi}^4 \,,
\eeqs
where  the superpotential is given by~\cite{DeWolfe:1999cp}
\beqs
{\mathcal W}_6(\phi) & \equiv & -2 - \frac{\Delta}{2} {\phi}^2\,.
\label{Eq:super}
\eeqs
The choice of a simple quadratic superpotential  provides a neat field-theory 
interpretation for backgrounds in which $\phi$ is non-zero: asymptotically in the UV, the dual field theory flows towards a CFT in $D-1=5$ dimensions,
deformed by the insertion of an operator ${\cal O}$ with scaling dimension given by ${\rm max}(\Delta,5-\Delta)$, and the two parameters appearing in the solution of the corresponding second-order classical 
equations correspond in field-theory terms to the coupling and condensate associated with ${\cal O}$.

We treat  $\Delta$ as a free parameter---in top-down models, by contrast, its counterpart descends from first principles.
In all three examples discussed in Refs.~\cite{Elander:2020ial,Elander:2020fmv,Elander:2021wkc},
the first-order equations involve choices of $\Delta > (D-1)/2$ (where $D$ is equal to $6$, $7$, and $5$, respectively, in those three models). Later in the paper, we will discuss the differences emerging for $0 < \Delta < (D-1)/2 = 5/2$.
For the time being, it suffices to notice that the counter-term used for holographic renormalisation
coincides with ${\mathcal W}_6$ when $\Delta<5/2$, while for $\Delta>5/2$ one needs 
the associated superpotential
\beqs
\overline{\mathcal W}_6
&=&-2-\frac{1}{2}(5-\Delta) {\phi}^2
-\frac{25 (2 \Delta -5)}{16 (4 \Delta -15)}{\phi}^4
-\frac{125 (2 \Delta -5) \left(4 \Delta ^2-15 \Delta +25\right)}{64 (4 \Delta -15)^2 
(6   \Delta -25)}{\phi}^6\,+\,\cdots\,,
\label{Eq:AW}
\eeqs
which is only known perturbatively in $\phi$, yet also solves Eq.~\eqref{eq:VfromW}.
In Appendix~\ref{Sec:super}, we show this superpotential expanded to higher orders.
Pathological values of $\Delta=\frac{15}{4}\,,\,\frac{25}{6}\,,\cdots\,,$ appear at increasing orders in $\phi$.
In the following, we will avoid using these special pathological values.
Another special case is $\Delta=5/2$, which requires a separate treatment, as we shall see.

The boundary terms in Eq.~(\ref{Eq:bbcc})
 play an important role in this paper.
They determine the boundary conditions
 for the classical background solutions, via the consistency requirements on the 
 variational principle. 
 They are used in the calculation of the spectrum of fluctuations around such background solutions.
And finally, they enter into the calculation of the free
 energy---but notice that this last calculation requires
a modification of the UV-boundary terms, in which $\lambda_2$ (and $f_2$) must be
 replaced by the
 counter-terms required for holographic renormalisation.

\subsection{Dimensional reduction to five dimensions}
\label{Sec:reduction}

One of the dimensions is compact, the coordinate $0 \leq \eta < 2\pi$
describing a circle. In reducing the action to five dimensions, we write the metric as
\begin{align}
	\dd s_6^2 &= e^{- 2\chi} \dd x_5^2 + e^{6\chi} \left(\dd \eta + \chi_M \dd x^M \right)^2 \,,
	\label{Eq:6}
\end{align}
where the space-time index is $M = 0,1,2,3,5$. The reduced action then becomes
\beqs
\label{eq:action5d}
	\mathcal S_5 &=& \mathcal S_5^{(bulk)} + \sum_{i=1,2} \mathcal S_{4,i} \,, \\
	\mathcal S_5^{(bulk)} &= &\int \dd^5 x \sqrt{-g_5} \, \BigCurlyLeft
	\frac{R}{4} - \frac{1}{2} g^{MN} \left[ 6 \partial_M \chi \partial_N \chi  \frac{}{}+\frac{}{} 
	\partial_M {\phi} \partial_N {\phi} \right] - e^{-2\chi} \mathcal V_6(|\phi|) 
	\nonumber\\
	&&\hspace{2.4cm}
 - \frac{1}{16} e^{8\chi} g^{MP} g^{NQ} F^{(\chi)}_{MN} F^{(\chi)}_{PQ} 
	\BigCurlyRight \,, \\
	\mathcal S_{4,i} &=& (-)^i \int \dd^4 x \sqrt{-\tilde g} \, \bigg\{ \frac{K}{2} + e^{-\chi} \lambda_i(\phi) + e^{-\chi} f_i(\chi) \bigg\} \bigg|_{\rho = \rho_i} \,,
\eeqs
where the five-dimensional metric $g_{MN}$ has determinant $g_5$, the induced 
metric on the boundaries is $\tilde g_{MN}$, the five-dimensional Ricci scalar is $R$, and $K$ is the extrinsic curvature. 
The field strength for the vector $\chi_M$ is given by $F^{(\chi)}_{MN} = \partial_M \chi_N - \partial_N \chi_M$. 
The functions $f_i$ of the six-dimensional theory depend explicitly on $\chi$, as is required in order 
to obtain solutions that lift to geometries in six dimensions in which
 the circle shrinks smoothly.

The scalars $\Phi^a=\{\phi,\chi\}$ describe a sigma-model coupled to gravity, the action of which is the same as in Eq.~\eqref{eq:sigmamodelaction} with $D=5$ and sigma-model metric
$G_{ab}={\rm diag}(1,\,6)$. We consider background solutions in which $\chi_M = 0$, while the metric $g_{MN}$, $\phi$, and $\chi$ depend on the radial coordinate only. The metric in five dimensions takes the domain-wall (DW) form
\beq
	\dd s_5^2 = \dd r^2 + e^{2A(r)} \dd x_{1,3}^2 = e^{2\chi(\rho)} \dd \rho^2 + e^{2A(\rho)} \dd x_{1,3}^2 \,,
\eeq
where we have
adopted the convenient choice of radial coordinate $\dd \rho = e^{-\chi} \dd r$. The  background fields satisfy the equations of motion
\begin{align}
	\label{Eq:1}
	\partial_\rho^2 \phi + (4 \partial_\rho A - \partial_\rho \chi) \partial_\rho \phi &= \frac{\partial \mathcal V_6}{\partial {\phi}} \,, \\
	\label{Eq:2}
	\partial_\rho^2 \chi + (4 \partial_\rho A - \partial_\rho \chi) \partial_\rho \chi &= - \frac{\mathcal V_6}{3} \,, \\
	\label{Eq:3}
	3 (\partial_\rho A)^2 - \frac{1}{2} \left(\partial_\rho \phi\right)^2 
	- 3 (\partial_\rho \chi)^2 &= - \mathcal V_6 \,,
\end{align}
with boundary conditions given by
\begin{align}
\label{eq:BCX}
	\left( \partial_\rho \phi - \frac{\partial \lambda_i}{\partial \phi} \right) \bigg|_{\rho_i} &= 0 \,, \qquad
	\left( 6 \partial_\rho \chi + \lambda_i + f_i - \frac{\partial f_i}{\partial \chi} \right) \bigg|_{\rho_i} = 0 \,, \qquad
	\left( \frac{3}{2} \partial_\rho A + \lambda_i + f_i \right) \bigg|_{\rho_i} = 0 \,.
\end{align}
For vanishing $f_i = 0$, one obtains solutions that lift to domain walls in six dimensions.

The solutions satisfy the following equation:

\beq
0=
12(\partial_{\rho} { A})^2\,+\,
{3}\partial_{\rho}^2{A}\,-\,3\partial_{\rho}\chi\partial_{\rho}A\,+\,4 {\mathcal V}_6
\,,
\eeq
which can be combined with Eq.~(\ref{Eq:2}) to yield the conservation law (see also Ref.~\cite{Elander:2020ial})
\beq
\label{Eq:conservation}
\partial_{\rho}\left[\frac{}{}e^{4A-\chi}\left(\partial_{\rho}A-4\partial_{\rho}\chi\right)\right]=0\,.
\eeq
This defines a conserved quantity, which vanishes for DW solutions in six dimensions, 
for which the metric $\dd s_6^2$ in Eq.~(\ref{Eq:6}) has the warp factor defined by ${\cal A} \equiv A-\chi\,=\,3\chi$ (or $A=4\chi$).

\section{Classes of solutions}
\label{Sec:solutions}

Here, we introduce the three main classes of solutions that we study in the following.
We explain the naming choices, before
we exhibit the functional form of the solutions,
because they introduce some mild abuse of language, driven mostly by analogy,
which we want to alert the reader to.
We start from solutions to the first-order equations derived from  the superpotential formalism,
along the lines of fake supergravity~\cite{Freedman:2003ax}.
We call them supersymmetric solutions, despite the fact that there is no supersymmetry---the theory is bosonic. 
The reason for this naming convention is an analogy with top-down models derived from higher-dimensional 
supergravity, for which
the first-order equations coincide with the Bogomol'nyi--Prasad--Sommerfield (BPS) constraints.

The most important class of solutions of interest, and for which we compute the spectrum
of fluctuations, later in the paper, is denoted as
confining solutions, again with abuse of language. In models in which a 
lift to a higher dimensional supergravity derived from a string theory exists, it might be
possible to compute the Wilson loop, along the lines of other holographic models~\cite{Rey:1998ik,
Maldacena:1998im,Kinar:1998vq,Brandhuber:1999jr,Avramis:2006nv,Nunez:2009da}, 
to expose the area-law expected from  confinement.
This 
being a bottom-up model, such calculation cannot be carried out,
yet the solutions are completely smooth and introduce a mass gap in the spectrum of fluctuations,
which are weaker requirements for the gravity dual of a confining theory.

Finally, we also introduce a set of singular backgrounds,
that we call domain-wall solutions, because in six dimensions they take the form of Poincar\'e domain walls.
There is again abuse of language: strictly speaking the singularity signifies that they should not be taken literally as background solutions, and used with caution---but see Ref.~\cite{Gubser:2000nd}. 
Our reason for considering them is that, as we shall see,
 they teach us something important about the 
other classes of solutions and their stability.

\subsection{UV expansions}
\label{Sec:UV}

All the solutions we work with have the same asymptotic behaviour for 
large $\rho$: they approach the ${\phi}=0$ critical point of ${\cal V}_6$, with
$\chi\simeq\frac{1}{3}\rho$
and $A\simeq\frac{4}{3}\rho$
 (equivalently, ${\mathcal A} \simeq \rho$). Asymptotically in the UV, the gravity duals have common interpretations in terms of (relevant or marginal) deformations
of the same five-dimensional CFT.
We hence
classify all the solutions in terms of a power expansion
in the small parameter $z \equiv e^{-\rho}$.
The expansion depends on five free parameters.

Two parameters are additive contributions to $\chi$ and $A$,
which we denote $\chi_U$ and $A_U$, respectively. 
One free parameter appears in $\chi$ in the coefficient of the $z^5$ term, and we denote it as $\chi_5$. 
We fix an additive contribution to $\chi_5$ so that if $A=4\chi$ (as is the case for the DW solutions), then $\chi_5=0$.
For generic values of $\Delta$, 
 the remaining two free parameters  appear in the expansion of $\phi$ at orders $z^{\Delta}$ and 
$z^{5-\Delta}$, and  we call $\phi_J$ ($\phi_V$) the coefficient in front of the 
term with the smallest (largest) exponent $\Delta_J$ ($\Delta_V$).
The expansions take the generic form:
\begin{align}
	\phi(z) &= \phi_{J} z^{\Delta_J}\,+\,\cdots\, + \phi_{V} z^{\Delta_V} \,+\,\cdots \,, \\
	\chi(z) &= \chi_U - \frac{1}{3} \log(z)  \,+\,\cdots + ({\chi_5}+\cdots) z^5  \,+\,\cdots \,, \\
	A(z) &= A_U - \frac{4}{3} \log(z)  \,+\,\cdots \,.
\end{align}
In the special case with $\Delta=5/2$, for which the two
independent parameters appear in front of the $z^{5/2}$ and
$z^{5/2}\log(z)$ terms in the expansion of $\phi$,
we denote the former as $\phi_V$, and the latter as $\phi_J$.
As we shall see, these coefficients are very important in determining the physical properties 
of the solutions and their field-theory duals. For example, the
 free energy depends on a combination of $\phi_{J} \phi_{V}$ and $\chi_5$. We further note that, without loss of generality, one may restrict attention to solutions for which 
$A_U = 0 = \chi_U$.

The expansion depends non-trivially on $\Delta$. For definiteness, we report here the case $\Delta = 3$, while more examples can be found in Appendix~\ref{Sec:UVsolutions}: 
\begin{align}
	\phi(z) &= \phi_J z^2 + \phi_V z^3 - \frac{25}{48} \phi_J^3 z^6 
	-\frac{57}{80} \phi_J^2 \phi_V z^7
	+ \mathcal O\left( z^8 \right) \,, \\
	\chi(z) &= \chi_U - \frac{1}{3} \log(z) - \frac{1}{24} \phi_J^2 z^4 + 
	\left(\chi_5 -\frac{2}{25}\phi_J\phi_V\right)z^5 - \frac{1}{24} \phi_V^2 z^6 + \mathcal O\left( z^8 \right) \,, \\
	A(z) &= A_U - \frac{4}{3} \log(z) - \frac{1}{6} \phi_J^2 z^4 +
	 \left( \frac{1}{4} \chi_5 - \frac{8}{25} \phi_J \phi_V \right) z^5 - \frac{1}{6} \phi_V^2 z^6 + \mathcal O\left( z^8 \right) \,.
\end{align}

\subsection{Supersymmetric solutions}
\label{Sec:supersymmetric}

As the scalar potential $\mathcal V_6$ is written in terms of a superpotential ${\mathcal W}_6$, 
a special class of  six-dimensional DW  
solutions
(for which $A(\rho) = 4 \chi(\rho) = \frac{4}{3} \mathcal A(\rho)$) is recovered 
 by solving the first-order equations
\beqs
\partial_{\rho} {\cal A}&=&-\frac{1}{2}{{\cal W}_6}\,=\,1+\frac{\Delta}{4} \phi^2
\,,\\
\partial_{\rho} \phi &=&\frac{\partial{{\cal W}_6}}{\partial \phi} \,=\,-\Delta \phi
\,,
\eeqs
which, as anticipated, we call supersymmetric solutions:
\beqs
	\phi(\rho) &=& \phi_c \, e^{-\Delta \rho}\,=\,\phi_c\,z^{\Delta} \,, \\
	 \mathcal A(\rho) &=& \rho - \frac{1}{8} \phi_c^2 \, e^{-2\Delta \rho} 
	 \,=\,- \log(z) -\frac{1}{8}\phi_c^2\,z^{2\Delta}\,.
\eeqs
Besides $\phi_c$, a second, additive integration constant has been omitted from ${\mathcal A}$. 
We recover the AdS$_6$ geometry with $\phi_c = 0$.

\subsection{Confining solutions}
\label{Sec:confinement}

The aforementioned confining solutions are such that
 the circle parametrised by $\eta$ shrinks to zero size at some point $\rho_o$ of the radial direction $\r$---which 
 is hence bounded from below as 
$\rho_o<\rho_1\leq \rho < \r_2\rightarrow +\infty$---and are also
completely regular and smooth, as the metric in six dimensions has
 finite curvature invariants and there is no conical singularity.
By power expanding  near the end of space, for small $(\rho - \rho_o)$, we find that such solutions have the following form
\begin{align}
\label{Eq:IR1}
	\phi(\rho) &= \phi_I - \frac{1}{16} \Delta \phi_I \left( 20 + \Delta \left( 5 \phi_I^2 - 4 \right) \right) (\rho - \rho_o)^2 + \mathcal O\left((\rho - \rho_o)^4\right) \,, \\
	\label{Eq:IR2}
	\chi(\rho) &= \chi_I + \frac{1}{3} \log(\rho - \rho_o) + \frac{1}{288} \left( -80 + 8 \left( \Delta - 5 \right) \Delta \phi_I^2 - 5 \Delta^2 \phi_I^4 \right) (\rho - \rho_o)^2 + O\left((\rho - \rho_o)^4\right) \,, \\
	\label{Eq:IR3}
	A(\rho) &= A_I + \frac{1}{3} \log(\rho - \rho_o) + \frac{7}{576} \left( 80 +\Delta \phi_I^2 \left( 40 + \Delta \left( 5 \phi_I^2 - 8 \right) \right) \right) (\rho - \rho_o)^2 + \mathcal O\left((\rho - \rho_o)^4\right) \,,
\end{align}
where $\phi_I$, $\chi_I$, $A_I$, and $\rho_o$ are integration constants.

 The induced metric on the $(\rho,\eta)$ sub-manifold is
 \beqs
 \dd s^2_2 &=&\dd \rho^2 \,+\,e^{6\chi}\dd \eta^2 \,\simeq\,\dd \rho^2 \,+\,e^{6\chi_I}(\rho-\rho_o)^2\dd \eta^2\,+\,\cdots\,,
 \eeqs
 which is the metric of the plane, provided  $\eta$ has periodicity $2\pi$ and we fix $\chi_I = 0$,
to  avoid a conical singularity.

 As a side remark,
 when $\Delta=0$,
we can write the solution in closed form:
 \beqs
 \phi(\rho) &=& \phi_I 
 \,,\\
 \chi(\rho) &=&
  \chi_o -\frac{1}{5}\log\left[\cosh\left(\frac{5}{2}(\r-\r_o)\right)\right]+\frac{1}{3}\log\left[\sinh\left(\frac{5}{2}(\r-\r_o)\right)\right]
 \,,\\
A(\rho) &=&
{
 A_o +\frac{4}{15}\log\left[\sinh\left(5(\r-\r_o)\right)\right] + \frac{1}{15}\log\left[\tanh\left(\frac{5}{2}(\r-\r_o)\right)\right]
}
 \,,
 \eeqs
where $\chi_o$ and $A_o$ are additive integration constants.

 The  curvature invariants in six dimensions---the Ricci scalar $ \mathcal R \equiv { \mathcal R}_6$, the Ricci tensor squared $ \mathcal R^2_2\equiv { \mathcal R_6}_{\hat{M}\hat{N}}{ \mathcal R_6}^{\hat{M}\hat{N}}$, 
and the Riemann tensor squared  
$\mathcal R^2_{4}\equiv { \mathcal R_6}_{\hat{M}\hat{N}\hat{R}\hat{S}}{ \mathcal R_6}^{\hat{M}\hat{N}\hat{R}\hat{S}}$---can
be written in terms of the non-trivial quantity
\beqs
d&\equiv& A-4\chi\,,
\eeqs
which vanishes on the DW solutions.
After using the equations of motion, we find the following~\cite{Elander:2020ial}:
\beqs
\label{eq:curvatureinvariants}
\mathcal R &=& 6\mathcal{V}_{6}+2\big(\partial_{\rho}{\phi}\big)^{2} \,,  \\
\mathcal R^{2}_{2} &=& 6\mathcal{V}_{6}^{2}+4\mathcal{V}_{6}\big(\partial_{\rho}{\phi}\big)^2
+4\big(\partial_{\rho}{\phi}\big)^{4} \,, \\
\mathcal R^{2}_{4}
&=&\frac{1}{250} \Bigg(32(\partial_{\rho}d)^2 \Big(4\partial_{\rho}d \sqrt{36(\partial_{\rho}d)^2+15\sqrt{5}
	\sqrt{6 \mathcal R^{2}_{ 2}- \mathcal R^{2}}-30
	\mathcal R}+24(\partial_{\rho}d)^2 \nonumber \\ 
&&\hspace{30mm}+5 \sqrt{5} \sqrt{6 \mathcal R^{2}_{ 2}- \mathcal R^{2}}-10 \mathcal R\Big)-25
\left( \mathcal R^{2}-10 \mathcal R^{2}_{ 2}\right)\Bigg) \,.
\eeqs

All of these quantities are finite when $\rho\rightarrow \rho_o$, as can be seen by explicitly using the IR expansions:
\beqs
\lim_{\r\rightarrow \r_o} \mathcal R &=&
 -30 - {3\Delta (5 - \Delta)} {\phi_I}^2 - \frac{15 \Delta^2}{8} {\phi_I}^4
\,,\\
\lim_{\r\rightarrow \r_o} \mathcal R^{2}_{2} &=& \frac{1}{6} \left(\lim_{\r\rightarrow \r_o} \mathcal R \right)^2
\,,\\
\lim_{\r\rightarrow \r_o} \mathcal R^{2}_{4}
&=& \frac{1}{3} \left(\lim_{\r\rightarrow \r_o} \mathcal R \right)^2
\,.
\eeqs
Interestingly though, as long as $\Delta \neq 0$, all these invariants diverge for $\phi_I\rightarrow \infty$,
suggesting a priori that we should not be allowed to take the value of $\phi_I$ to be arbitrarily large.

\subsection{Singular domain-wall solutions}
\label{Sec:DW}

The last class of solutions of interest to this paper is given by singular DW solutions.
They obey the DW ansatz $A=4\chi=\frac{4}{3}{\cal A}$. Their IR expansion depends explicitly on $\Delta$,
and reads as follows:
\beqs
\label{Eq:IR4}
\phi(\rho)&=&\phi_I-\sqrt{\frac{2}{5}} \log (\rho-\r_o ) 
+\frac{(\rho-\r_o)^2 }{25920}\nonumber
\left(2 \sqrt{10} (6 \Delta  \log (\rho-\r_o ) (3 \log (\rho-\r_o ) (\Delta  \log
   (\rho-\r_o ) (3 \log (\rho-\r_o )+2)
   \right.\\
   &&\left.\nonumber
   -23 \Delta +60)+37 \Delta +60)   -\Delta  (47 \Delta
   +660)+5400)\right.\\
   &&\left.
+15 \Delta  \phi_I \left(-60 \phi _I^2 (6 \Delta  \log (\rho-\r_o )+\Delta )+6
   \sqrt{10} \phi_I(6 \Delta  \log (\rho-\r_o ) (3 \log (\rho-\r_o )+1)\right.\right.\\
   &&\left.\left.\nonumber
   -23 \Delta +60)-4 (6 \log
   (\rho-\r_o ) (3 \Delta  \log (\rho-\r_o ) (2 \log (\rho-\r_o )+1)\right.\right.\\
   &&\left.\left.
-23 \Delta +60)+37 \Delta +60)+45
   \sqrt{10} \Delta  \phi_I^3\right)\right)  \,+\,{\cal O}((\r-\r_o)^4)\nonumber
\,,\\
\label{Eq:IR5}
{\cal A}(\rho)&=&
\frac{1}{5}\log(\r-\r_o)\nonumber
+\frac{(\rho-\r_o)^2 }{12960}
\left(2 (6 \Delta  \log (\rho-\r_o ) (3 \log (\rho-\r_o )   (\Delta  \log (\rho-\r_o ) (3 \log (\rho-\r_o )-10)
\right.\\
   &&\left.
+7 \Delta +60)-5 (\Delta +60))
+(1140-17
   \Delta ) \Delta +5400)\nonumber
   \right.\\
   &&\left.
   +3 \Delta  \phi_I \left(2 \sqrt{10} (5 (\Delta +60)
   -6 \log
   (\rho-\r_o ) (3 \Delta  \log (\rho-\r_o ) (2 \log (\rho-\r_o )-5)+7 \Delta +60))
 \right.  \right.\\
   &&\left. \left. \nonumber
   +15 \phi_I \left(2
   (6 \Delta  \log (\rho-\r_o ) (3 \log (\rho-\r_o )-5)+7 \Delta +60)
\right.  \right. \right.\\
   &&\left.\left.\left.
   +\Delta  \phi_I \left(2
   \sqrt{10} (5-6 \log (\rho-\r_o ))+15 \phi_I\right)\right)\right)\right)
   \,+\,{\cal O}((\r-\r_o)^4)\nonumber
\,.
\eeqs
In these expressions, $\phi_I$ is an integration constant, and without loss of generality we omitted an additive integration constant in $\mathcal A$.
We recall that  the system of equations is symmetric under the change $\phi\rightarrow -\phi$,
hence a second branch of solutions can be obtained by just changing the sign of $\phi$.

These solutions are singular at the end of space.
Their nature in the gravity theory is unclear, and they do not admit a transparent field theory interpretation.
In particular, we do not compute the spectrum of their fluctuations---otherwise interpreted in terms
of bound states in a putative dual theory---as there is no clear sense in which this would provide 
information about observable quantities.
Yet, we do compute their free energy later in the paper, because, as we shall see, 
for some choices of parameters, solutions of this class are energetically favoured
over the confining ones.
We will come back to this discussion in due time; we only anticipate here
that this finding will force us to restrict our physics discussions of the confining solutions to a subregion of the full parameter space.

\section{Mass spectrum of fluctuations}
\label{Sec:glueballs}

In this section, we restrict our attention to the confining solutions only.
Given a background solution in the gravity theory,  one can linearise the equations of motion of  small fluctuations
 around it.
The resulting mass spectra of fluctuations of the bosonic fields
correspond, following  the  {gauge-gravity} duality dictionary,
 to spin-$0,1,2$ composite particles in  the (putative) dual {confining} field theory in four 
dimensions. 
To perform the calculations, we adopt the gauge-invariant formalism developed in 
Refs.~\cite{Bianchi:2003ug,Berg:2005pd,Berg:2006xy,Elander:2009bm,Elander:2010wd}. 
We start from the scalar fluctuations $\mathfrak{a}^{a}=\mathfrak{a}^{a}(q,\rho)$, where $q^\mu$ is the four-momentum, which  obey the following equations
(see also Appendix~\ref{Sec:Afluctutations}):
\beqs
\label{eq:scalarflucs1}
{\Big[\partial^2_{\rho} + (4\partial_{\rho}A - \partial_\rho \chi)\partial_{\rho}
-e^{2\chi-2A}q^2 \Big]}\mathfrak{a}^{a} - e^{2\chi} \mathcal X^{a}_{\ c}
\mathfrak{a}^{c}&=&0\,.
\eeqs
Because the sigma-model metric is simply $G_{ab}={\rm diag}(1,\,6)$, 
   in the basis $\Phi^a=\{ \phi,\chi \}$,
   the sigma-model connection vanishes, greatly simplifying the equations of motion, 
   as $\mathcal X^{a}_{\ c}$ reads as follows:
   \beqs
\label{eq:scalarflucs2}
 \mathcal X^{a}_{\ c}&\equiv&
 \frac{\partial}{\partial \Phi^c}\bigg(G^{ab}\frac{\partial (e^{-2\chi}\mathcal{V}_6)}{\partial  \Phi^{b}}\bigg)\, 
+\frac{4}{3\partial_{\rho} A}
\bigg[\partial_{\rho} \Phi^{a}\frac{\partial (e^{-2\chi}\mathcal{V}_6)}{\partial  \Phi^{c}}+G^{ab}\frac{\partial (e^{-2\chi}\mathcal{V}_6)}{\partial  \Phi^{b}}\partial_{\rho}\Phi^{d}G_{dc}\bigg]\nonumber\\
&&
+\frac{16(e^{-2\chi}\mathcal{V}_6)}{9(\partial_{\rho}A)^{2}}\partial_{\rho} \Phi^{a}\partial_{\rho} \Phi^{b}G_{bc}\, .
\eeqs
In all these expressions, the functions  $A$,  $\Phi^a=\{\phi,\chi\}$, and ${\mathcal V}_6$
 are evaluated on the background.

A discrete spectrum can be found by imposing the following boundary conditions:\footnote{The equivalent form of the boundary conditions given in Eq.~(14) of Ref.~\cite{Elander:2014ola} is convenient in numerical implementations.}   
\be
	\label{eq:scalarflucs3}
	\left.\frac{}{}e^{-2\chi}\partial_{\rho} \Phi^{c}
	\partial_{\rho} \Phi^{d}G_{db}\partial_{\rho}\mathfrak{a}^{b}\right|_{\rho_{i}}
	=\left.\left[\frac{}{}\frac{3\partial_{\rho}A}{2}e^{-2A}q^{2}\delta^{c}_{\ b}
	+\partial_{\rho} \Phi^{c}\bigg(\frac{4\mathcal{V}_6}{3\partial_{\rho}A}\partial_{\rho} \Phi^{d}G_{db} + \frac{\partial\mathcal{V}_6}{\partial\Phi^{b}} \bigg) \right]\mathfrak{a}^{b}\right|_{\rho_{i}}\,,
\ee
Physical composite states in the dual theory have mass  $M^2=-q^2$, corresponding to choices 
of $q^2$ for which the boundary conditions at the two boundaries can be satisfied simultaneously.

The equations of motion for  the (transverse and traceless) tensor fluctuations $\mathfrak e^{\mu}_{\ \nu}$ are the following~\cite{Elander:2018aub}:
\beqs
0&=&\left[\frac{}{}\partial^{2}_{\r} + (4\partial_{\r}A-\partial_{\r}\chi) \partial_{\r}-e^{2\chi-2A}q^2 \right]\mathfrak{e}^{\mu}_{\ \nu}\,,
\eeqs
while for the vector $\chi_M$ one  looks at the gauge-invariant transverse polarisations, which obey the bulk equation
\beqs
0&=& P^{\mu\nu}\left[\frac{}{} \partial^2_{\r} + ( 2\partial_{\r} A + 7 \partial_{\r} \chi )\partial_{\r} - e^{2\chi-2A} q^2 \right] \chi_{\nu} \,,
\eeqs
where $P^{\mu\nu} \equiv \eta^{\mu\nu} - \frac{q^\mu q^\nu}{q^2}$. To compute the  
spectrum of masses, $M > 0$, of the bound states in the dual confining theory, one can impose Neumann boundary conditions on these two kinds of fluctuations.

The physical results can be recovered by first computing the spectrum for finite values of $\rho_1$ and $\rho_2$, 
and then by repeating the calculations to extrapolate toward the limits $\r_1 \rightarrow \r_o$ 
and $\r_2 \rightarrow +\infty$~\cite{Elander:2010wd,Elander:2013jqa,Elander:2018aub}. 
Equivalent results can be obtained with an alternative numerical strategy,
 which improves the convergence in such limits by making use of the 
UV and IR expansions of the fluctuations. One decomposes them in dominant and subdominant modes, 
notices that the boundary conditions  are equivalent to suppressing the dominant fluctuations, and 
then matches the solutions to the expansions, evaluated at finite $\r_{1,2}$.
This non-trivial process requires knowing the appropriate expansions
 at high orders in the small parameters ($z$ in the UV, and $\r-\r_o$ in the IR), but has the great advantage that one does not
 require extending the {background solutions of the non-linear equations of motion} to numerically challenging values  at large (small) $\rho-\rho_o$.
 This strategy was  successful in the study of the Klebanov-Strassler system~\cite{Berg:2006xy} and its baryonic branch~\cite{Elander:2017cle,Elander:2017hyr}, and we adopt it here. We report IR expansions, and one example of UV expansions, for the fluctuations in Appendix~\ref{sec:IRUVexpansions}.

As explained and exemplified in Ref.~\cite{Elander:2020csd}, as well as Appendix~\ref{sec:probedetail},
 the probe approximation is equivalent to ignoring the coupling of the states to the trace of the energy-momentum tensor,
 and hence discarding  the mixing in the physical states with the dilaton field.
This approximation fails to reproduce the mass of states
with substantial overlap with the dilatation operator.
When such states are light, they should hence be interpreted as approximate dilatons, turning this approximation
into a diagnostic tool for the identification of approximate dilatons. We will perform the analysis of the mass spectrum in the probe approximation only for the choice of $\Delta=5/2$, which turns out to be the most interesting in the context of this paper.

\begin{figure}[th]
\begin{center}
\includegraphics[width=16cm]{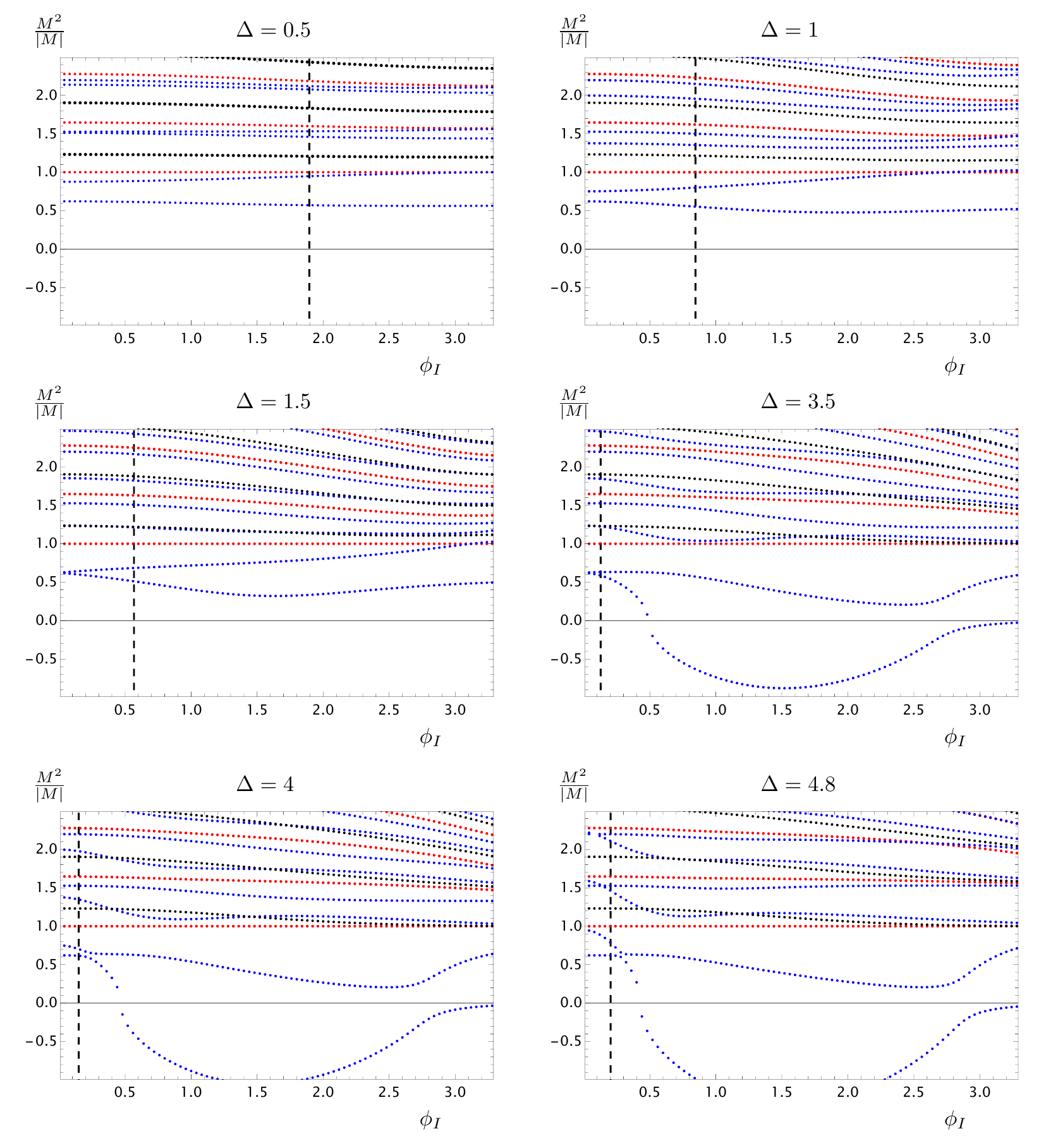}
\caption{Mass spectrum $\frac{M^2}{|M|}$ of fluctuations, computed for confining backgrounds, with various choices of $\Delta$, 
as a function of the IR parameter $\phi_I$.
For each $\Delta$, we show the spectrum of spin-0 (blue),
spin-1 (black), and spin-2 (red) states. The values of the IR and UV cutoffs in the calculations are respectively given by $\rho_1-\rho_o=10^{-9}$ and  $\rho_2-\rho_o=5$.
All masses are normalised to the mass of the lightest spin-2 state. 
The vertical dashed lines denote the critical value $\phi_I(c)$.}
\label{Fig:mass1}
\end{center}
\end{figure}

\begin{figure}[th]
\begin{center}
\includegraphics[width=16cm]{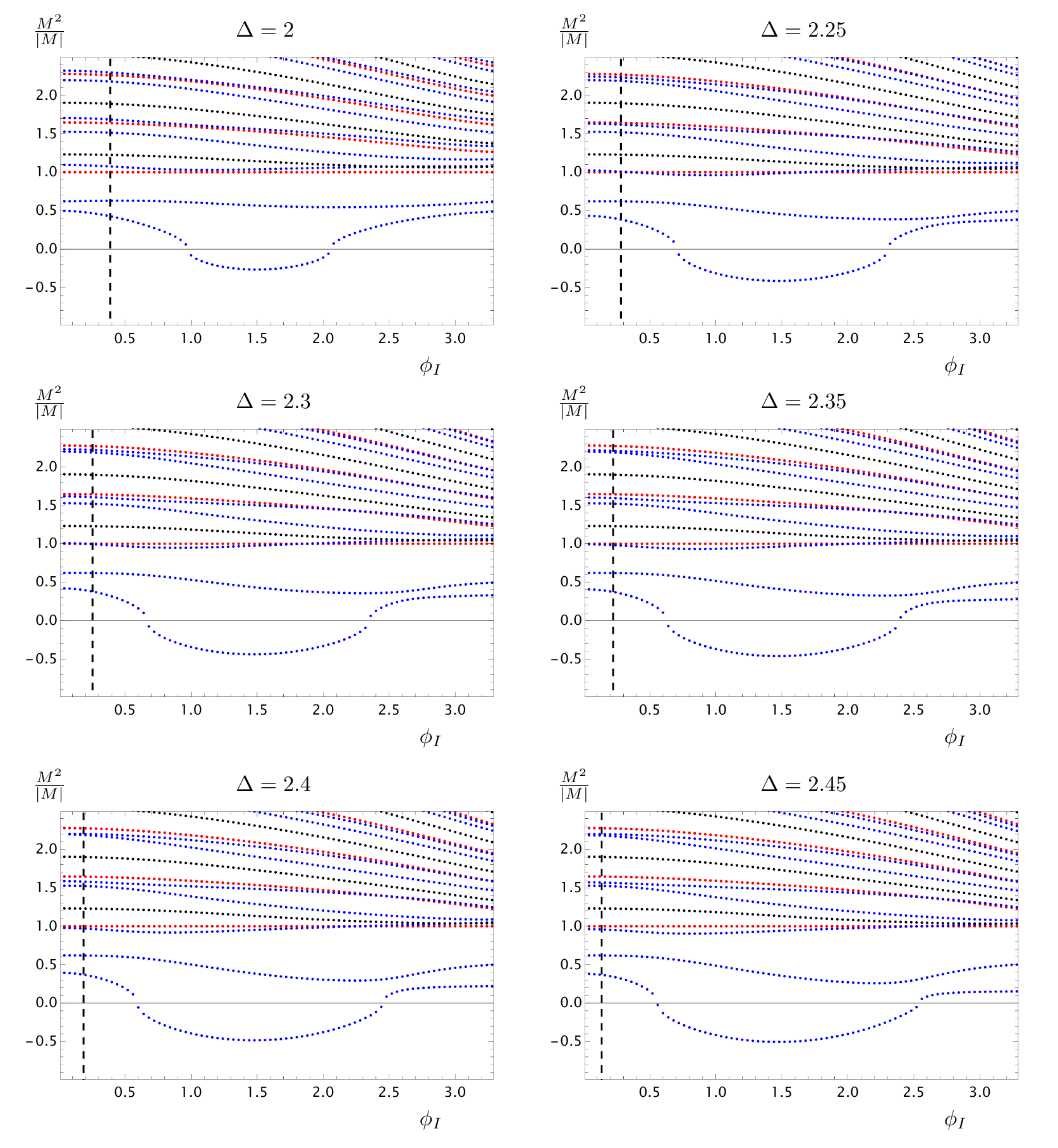}
\caption{Mass spectrum $\frac{M^2}{|M|}$ of fluctuations, computed for confining backgrounds, with various choices of $\Delta$, 
as a function of the IR parameter $\phi_I$.
For each $\Delta$, we show the spectrum of spin-0 (blue),
spin-1 (black), and spin-2 (red) states. The values of the IR and UV cutoffs in the calculations are respectively given by $\rho_1-\rho_o=10^{-9}$ and  $\rho_2-\rho_o=5$.
All masses are normalised to the mass of the lightest spin-2 state. The vertical dashed lines denote the critical value $\phi_I(c)$.}
\label{Fig:mass2}
\end{center}
\end{figure}

\begin{figure}[th]
\begin{center}
\includegraphics[width=16cm]{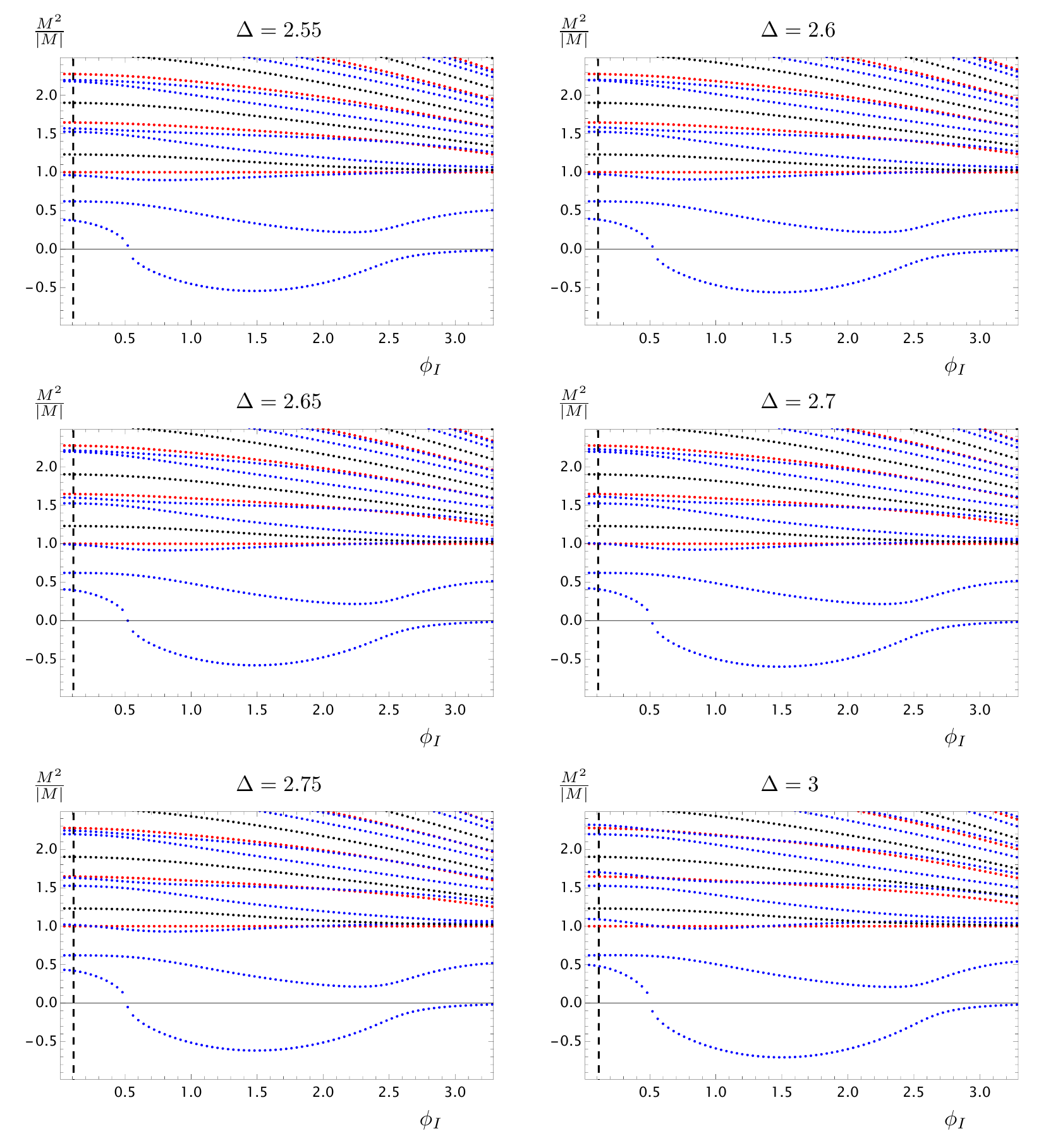}
\caption{Mass spectrum $\frac{M^2}{|M|}$ of fluctuations, computed for confining backgrounds, with various choices of $\Delta$, 
as a function of the IR parameter $\phi_I$.
For each $\Delta$, we show the spectrum of spin-0 (blue),
spin-1 (black), and spin-2 (red) states. The values of the IR and UV cutoffs in the calculations are respectively given by $\rho_1-\rho_o=10^{-9}$ and  $\rho_2-\rho_o=5$.
All masses are normalised to the mass of the lightest spin-2 state.
The vertical dashed lines denote the critical value $\phi_I(c)$.}
\label{Fig:mass3}
\end{center}
\end{figure}

\begin{figure}[th]
\begin{center}
\includegraphics[width=16cm]{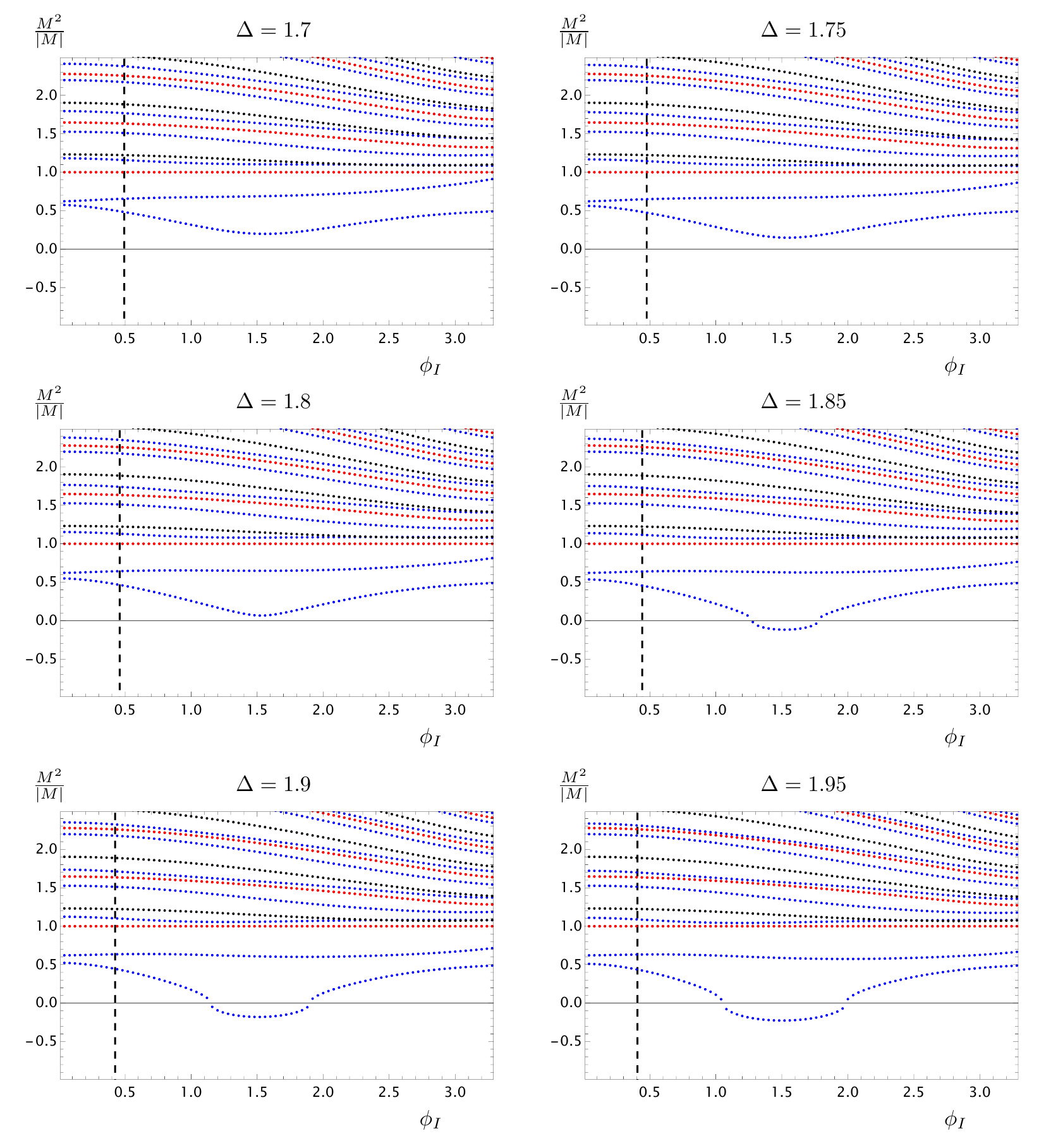}
\caption{Mass spectrum  $\frac{M^2}{|M|}$ of fluctuations, computed for 
confining backgrounds, with various choices of $\Delta$, 
as a function of the IR parameter $\phi_I$.
For each $\Delta$, we show the spectrum of spin-0 (blue),
spin-1 (black), and spin-2 (red) states. The values of the IR and UV cutoffs in the calculations are respectively given by $\rho_1-\rho_o=10^{-9}$ and  $\rho_2-\rho_o=5$.
All masses are normalised to the mass of the lightest spin-2 state.  
The vertical dashed lines denote the critical value $\phi_I(c)$.}
\label{Fig:mass4}
\end{center}
\end{figure}

The mass spectra are computed numerically,
and we report the results in Figs.~\ref{Fig:mass1},  \ref{Fig:mass2},  \ref{Fig:mass3},  and~\ref{Fig:mass4},
for a few significant choices of $\Delta$.
For each of them, we compute the spectrum of spin-0, spin-1, and spin-2 
fluctuations. 
We conveniently normalise the spectra so that the lightest tensor state has mass $M=1$.
For each value of $\Delta$, we check the convergence of the spectrum, as a function of $\r_1$ and $\r_2$
independently, and report numerical results obtained with choices for which the dependence on
the two cutoffs can be neglected in respect to  our numerical  accuracy goal of $0.5\%$.
Finally, we find it convenient to report in the plots also the critical 
value of  $\phi_I(c)$, obtained from the numerical study of the free energy, as detailed in the next section.

\section{Free energy}
\label{Sec:freeenergy}

The free energy density is given by the following expression:
\beqs
{\cal F}
&=&-\lim_{\rho_2\rightarrow +\infty}e^{4A-\chi}\Big(\frac{3}{2}\partial_{\rho}A+\mathcal{W}_{2}\Big)\Big|_{\rho_2}\,,
\label{Eq:Ffinal}
\eeqs
where ${\cal W}_2={\mathcal W}_6$ if $\Delta < 5/2$, or ${\cal W}_2=\overline{\mathcal W}_6$ if $\Delta>5/2$. 
This expression has been obtained 
 by adapting the results in Ref.~\cite{Elander:2020ial};  it
includes the contribution of the bulk action, evaluated by imposing the equations of motion, 
 the contribution of the boundary terms in the action, and the appropriate UV-localised counter-term
${\mathcal W}_2$, which is required by the rules of holographic renormalisation~\cite{Bianchi:2001kw,
Skenderis:2002wp,Papadimitriou:2004ap}, as it removes the UV divergences.
The conservation law
in Eq.~(\ref{Eq:conservation}) has been used  to evaluate at $\rho_2$ (the UV boundary) a term naturally 
defined at $\rho_1$ (the IR boundary).

For each choice of the parameter $\Delta$,
one expresses ${\cal F}$ in terms of 
 the coefficients appearing in the UV expansion of the background,
 which can be extracted by matching the expansions to numerical solutions.
Care must be taken to obtain results that converge as $\rho_2\rightarrow +\infty$.
We empirically  find that, for $\Delta<5/2$, the free energy density is
\beqs
{\cal F}&=&-\frac{1}{40}e^{4A_U-\chi_U}\left(16\Delta\left(\frac{5}{2}- \Delta\right)
\phi_{J}\phi_{V}\frac{}{}-75 \chi_5\right)\,,
\eeqs
while  for $\Delta>5/2$---and retaining in $\overline{\mathcal W}_6$ a sufficient number of terms to
ensure cancellation of all the divergences---the free energy density is
\beqs
{\cal F}&=&-\frac{1}{40}e^{4A_U-\chi_U}\left(16 (\Delta-5)\left(\frac{5}{2}-\Delta \right)\phi_{J}\phi_{V}\frac{}{}-75 \chi_5\right)\,.
\eeqs
We explicitly checked that this expression is accurate for all values of $\Delta$ considered in this paper.
In the special case in which the operator has dimension 
$\Delta = 5/2$, we find that 
\beqs
{\cal W}_2&=&-2-\frac{5}{4}\phi^2\left(1+\frac{2}{5\log(k\,z)}\right)\,,
\eeqs
and the free energy density is
\beqs
\mathcal F = \frac{1}{40} e^{4 A_U-\chi_U} \bigg( 20 \phi_{J} \phi_{V} - 4\phi_{J}^2 + 75 \chi_5 
- 20 \phi_{J}^2 \log(k) \bigg)\,.
\eeqs
The appearance of a logarithm introduces a residual scheme dependence, encoded in the parameter $k$. In the following, we make the choice $k = \Lambda$.

Following Refs.~\cite{Elander:2020ial,Elander:2020fmv,Elander:2021wkc},
we find it convenient to define a scale $\Lambda$ as follows~\cite{Csaki:2000cx}:
\begin{equation}
\label{Eq:Scale}
\Lambda^{-1}\equiv
\int_{\rho_o}^{\infty}\dd \rho\,e^{\chi(\rho)-A(\rho)}\,,
\end{equation}
with $\rho_o$ the end of space---other choices are admissible,
but this choice makes contact with earlier studies, where it has been shown that it allows to compare
the confining and DW classes of solutions.
We then express all of the quantities of interest in units of such scale $\Lambda$,
by defining the rescaled free energy density
\beqs
\hat {\mathcal F} &  \equiv& \frac{\mathcal F}{ \Lambda^5}\,,
\eeqs
and the rescaled source as
\beqs
\hat \phi_{J} &\equiv& \frac{\phi_{J}}{\Lambda^{\Delta_J}}\,,
\eeqs
where $\Delta_J$ is the dimension of the deforming coefficient (source)
associated with the dual operator of dimension $\Delta_V\equiv 5-\Delta_J$.
We hence define also the rescaled condensates to be
\beqs
\hat \phi_{V} &\equiv& \frac{\phi_{V}}{\Lambda^{\Delta_V}}\,,\\
\hat \chi_{5} &\equiv& \frac{\chi_{5}}{\Lambda^{5}}\,.
\eeqs

\begin{figure}[t]
\begin{center}
\includegraphics[width=16cm]{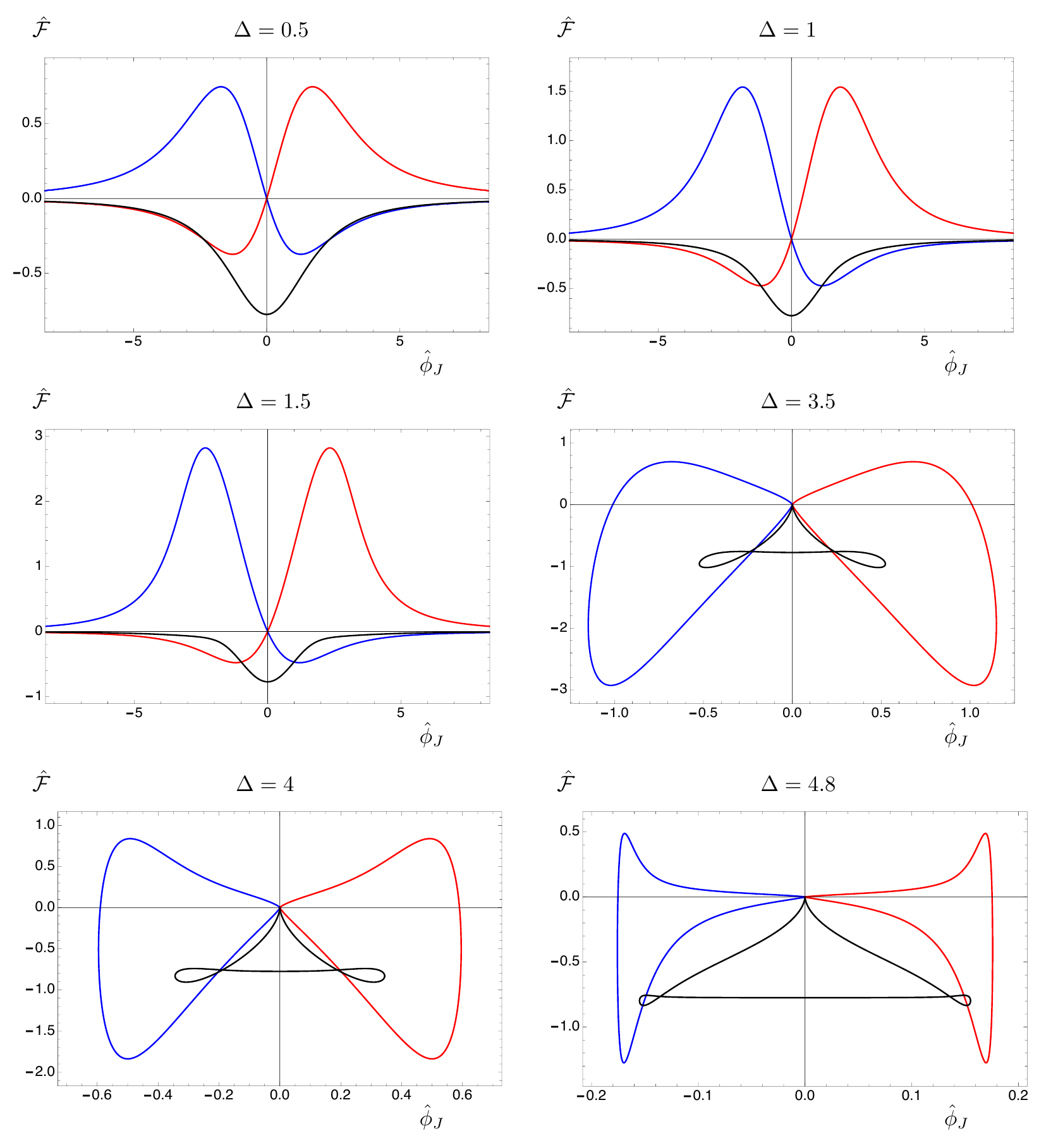}
\caption{The free energy density $\hat{\cal F}$ as a function of the source
$\hat{\phi}_{{J}}$,  in units of the scale $\Lambda$, for various choices of $\Delta$.
The black curve denotes the confining solutions, while the red and blue solutions are singular 
domain-wall ones.}
\label{Fig:FvsJ_1}
\end{center}
\end{figure}

\begin{figure}[t]
\begin{center}
\includegraphics[width=16cm]{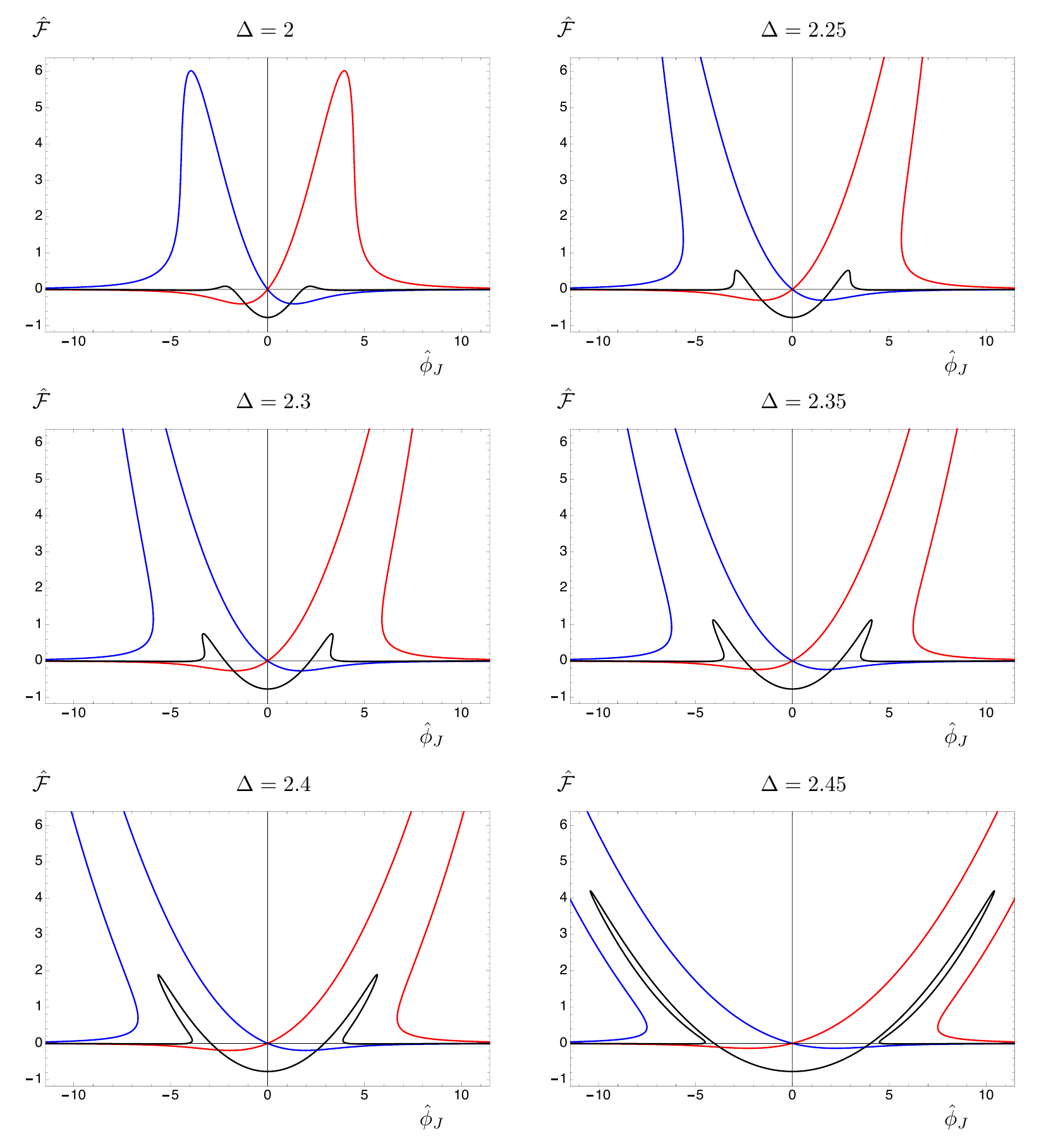}
\caption{The free energy density $\hat{\cal F}$ as a function of the source
$\hat{\phi}_{{J}}$,  in units of the scale $\Lambda$, for various choices of $\Delta$.
The black curve denotes the confining solutions, while the red and blue solutions are singular 
domain-wall ones.}
\label{Fig:FvsJ_2}
\end{center}
\end{figure}

\begin{figure}[t]
\begin{center}
\includegraphics[width=16cm]{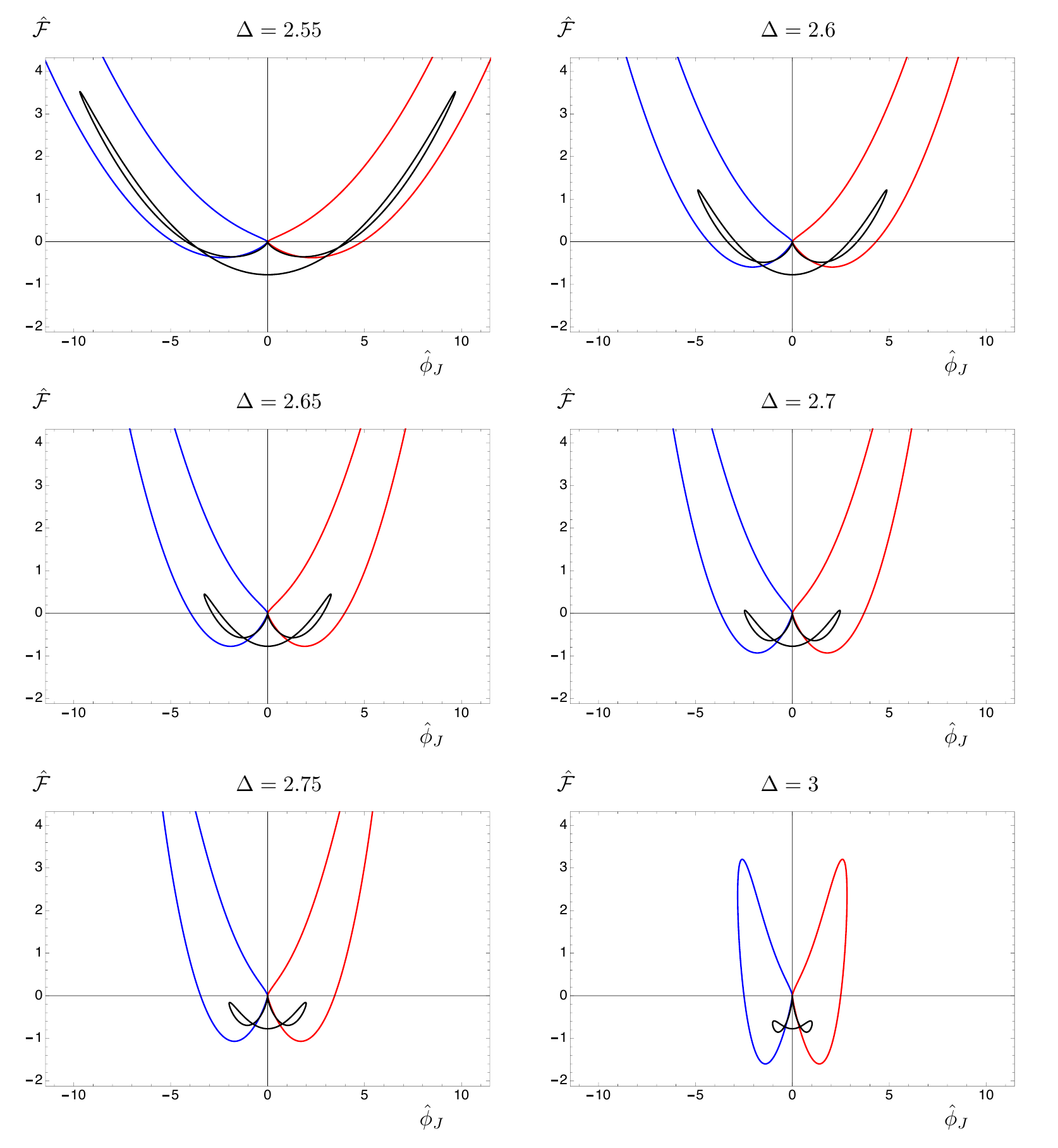}
\caption{The free energy density $\hat{\cal F}$ as a function of the source
$\hat{\phi}_{{J}}$,  in units of the scale $\Lambda$, for various choices of $\Delta$.
The black curve denotes the confining solutions, while the red and blue solutions are singular 
domain-wall ones.}
\label{Fig:FvsJ_3}
\end{center}
\end{figure}

\begin{figure}[t]
\begin{center}
\includegraphics[width=16cm]{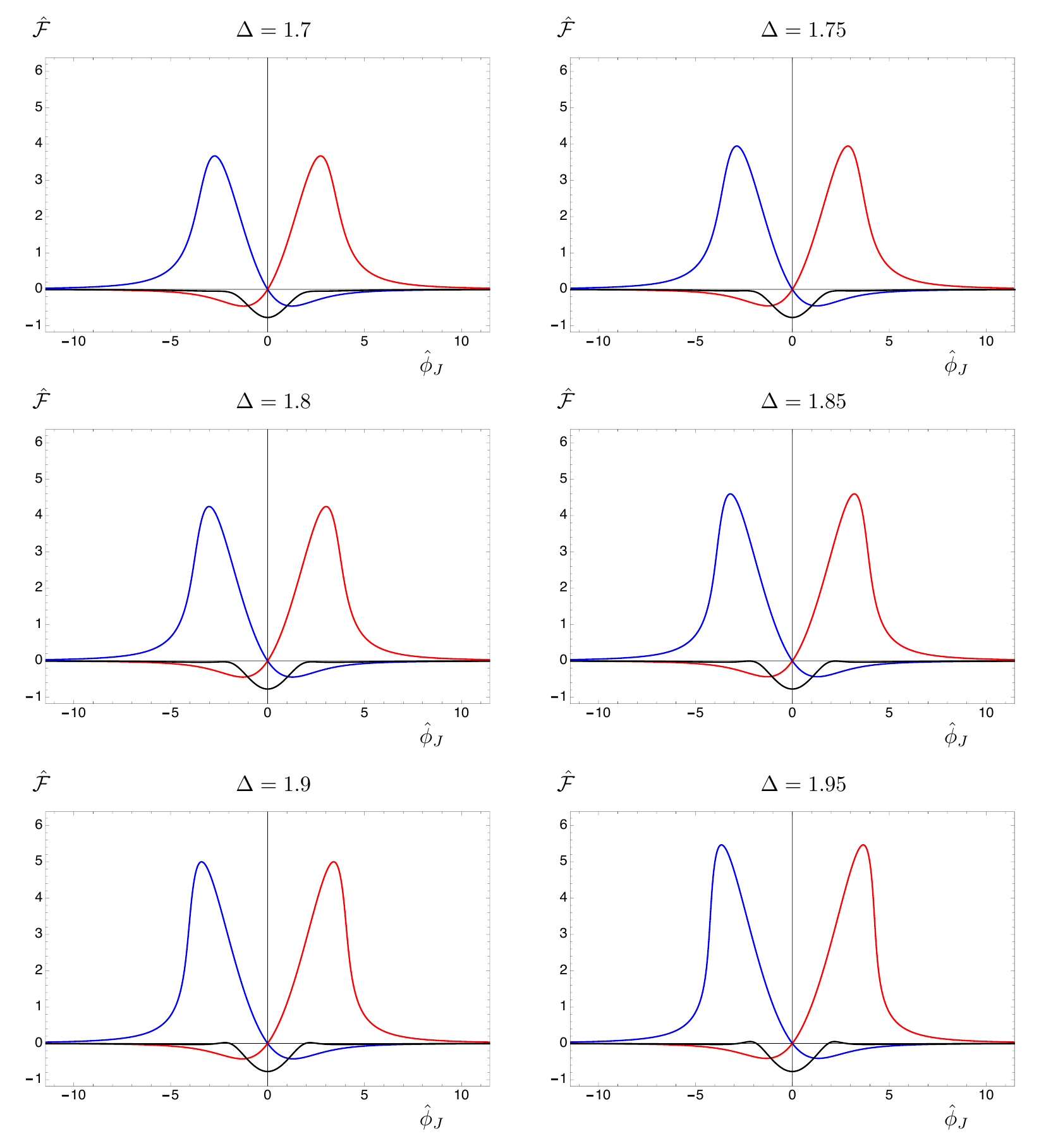}
\caption{The free energy density $\hat{\cal F}$ as a function of the source
$\hat{\phi}_{{J}}$,  in units of the scale $\Lambda$, for various choices of $\Delta$.
The black curve denotes the confining solutions, while the red and blue solutions are singular 
domain-wall ones.}
\label{Fig:FvsJ_4}
\end{center}
\end{figure}

Given a background solution, in order to compute the free energy
we proceed as follows.
First, we 
match it to the UV expansions, and determine coefficients $A_U$, $\chi_U$, $\phi_{J}$, $\phi_{V}$,
and $\chi_5$.  We then shift additively both the radial coordinate $\r$ and the definition of the function
$A(\r)$, to impose the constraints $A_U=0=\chi_U$. We repeat the determination of the
coefficients, and enter the results in the expression for the free energy density.
We compute the scale $\Lambda$ for each background---after the aforementioned shifts of $\r$ 
and $A(\r)$---and finally plot the resulting, rescaled quantities.
Examples are shown in Figs.~\ref{Fig:FvsJ_1}, \ref{Fig:FvsJ_2}, \ref{Fig:FvsJ_3}, and~\ref{Fig:FvsJ_4}.
We do so for both confining and DW singular solutions, and show them together for each  $\Delta$.

By inspection of the figures, we see the emergence of a
 common pattern; in all cases, the confining solutions minimise the free energy, $\hat {\mathcal{ F}}$,
for small choices of  the source $|\hat{\phi}_{J}|$, while the singular, DW solutions have the lowest $\hat{\cal F}$
for larger values of the source.
There is a critical value of  $|\hat{\phi}_{J}(c)|$, corresponding to a critical value of the
IR parameter $|\phi_I(c)|$, {where a first-order phase transition occurs. For $|\phi_I| \leq |\phi_I(c)|$, the confining solutions can be
physically realised, while for $|\phi_I| > |\phi_I(c)|$, the physical solution is unknown, so that we are forced to discard these latter regions of parameter space.
We are not claiming that the singular solutions are physically realised,
but simply that some other branch of solutions, 
not identified in this study, might exist, and take over the long-distance
dynamics in these cases.

When inspected  more closely, the figures show other interesting features.
For {$\Delta \geq 5/2$}, both confining and singular solutions cease to exist above some
value of $|\hat{\phi}_{J}({\rm max})|>|\hat{\phi}_{J}(c)|$. 
As we already know 
(from the aforementioned arguments)
that, at large $|\hat{\phi}_{J}|>|\hat{\phi}_{J}(c)|$,
another branch of solutions {must} exist (possibly in a more complete theory), this is not a particularly problematic finding. 
While this finding agrees with the results of the analysis of three distinct examples of top-down holographic models,
reported  in 
Refs.~\cite{Elander:2020ial,Elander:2020fmv,Elander:2021wkc},
it is interesting that this feature is absent for $\Delta<5/2$,
in which case confining and singular solutions exist for any choices of $\hat{\phi}_{J}$.

In thermodynamics, one can use  the concavity theorems to discuss the stability
of possible solutions, determined
 on the basis of the shape of the functional dependence of the free energy on the control parameters.
These arguments do not transfer directly to the case in question, 
because of the presence of divergences and scheme dependences
(see also Ref.~\cite{Bobev:2013cja}, for examples of similar considerations in a different context), and caution has to be exercised.
Yet, by comparing Figs.~\ref{Fig:FvsJ_1}, \ref{Fig:FvsJ_2}, \ref{Fig:FvsJ_3}, and~\ref{Fig:FvsJ_4}
with Figs.~\ref{Fig:mass1},  \ref{Fig:mass2},  \ref{Fig:mass3},  and~\ref{Fig:mass4}, one notices the following.
\begin{itemize}
\item A tachyon exists in the spectra of confining theories for values of $\Delta \gtrsim 1.8$,
 but only in a portion of parameter space (equivalently, for some range of $\phi_I$).
\item The portion of parameter space in which the tachyon appears
is always past the phase transition. 
\item There is a region of parameter space over which the mass of the lightest scalar is parameterically small,
near the point at which it turns tachyonic. 
But once more: this happens only past the phase transition,
in regions in which the 
confining solutions are, at best, metastable.
\end{itemize}

We performed an extensive study of the 
free energy for values of $\Delta \simeq 1.8$, nearby where the tachyon appears---see also the details in Appendix~\ref{sec:Fdetail}.
We find that the free energy of the confining backgrounds is a monotonic function for $\Delta$ below $\Delta \simeq 1.7$.
As expected, the relation between the behavior of ${\cal F}$ and the appearance a tachyon
is not precise, and furthermore these features always appear in a region of 
parameter space that has already been excluded, on the basis of earlier considerations.
We hence report on this feature, for completeness,  but do not explore it any further.

\section{Summary}
\label{Sec:summary}

We collect in Table~\ref{Fig:table} our numerical results: for  representative values of $\Delta$,
we tabulate  the critical parameters
$\hat{\phi}_{J}({\rm c})$ (source) and $\phi_I({\rm c})$ (IR asymptotic  value of $\phi$), 
the mass of the lightest scalar $M(c)$ (in units of the mass of the  lightest tensor),
in proximity of the transition. For completeness, we also list the value of the two parameters 
representing the
condensates in the dual field  theory: $\hat{\phi}_{V}(c)$ and $\hat{\chi}_5(c)$.
We show  in Fig.~\ref{Fig:phase}  the mass spectrum and free energy for the $\Delta=5/2$ case,
and in Fig.~\ref{Fig:summary} we display
the mass spectrum computed at the critical value $\phi_I(c)$, for each choice of $\Delta$.

The main results of the paper can be summarised as follows.
The model allows us to study the spectrum and the free energy for any generic value of the parameter $\Delta$,
which is equal
either to the dimensionality of the coupling deforming the dual CFT, or to the associated condensate.
Furthermore, for each $\Delta$ we restrict attention to regular backgrounds, the dual field theory of which 
has a mass gap and a discrete spectrum of bound states. 
These solutions of the gravity equations of motion in five dimensions  lift
to completely regular and smooth solutions in six dimensions, and  admit an end
of space in the radial direction $\r>\r_o$.
We compute the spectrum of fluctuations of solutions of this type, which with abuse of notation we call {\it confining}.
We find that for all values of the parameter $0<\Delta<5$ that we studied, there exists a first-order phase transition,
that bounds from above  the source associated with the scalar field $\phi$.
This is demonstrated by the existence of domain-wall solutions in six dimensions that, for large enough 
values of the deforming parameters, are energetically favoured (despite being singular) over the confining ones. 
We find  that for $\Delta \gtrsim 1.8$, the mass spectrum  contains a tachyon, provided the deforming parameter is large enough.
But this tachyon always 
appears only in an unphysical portion of parameter space, well beyond the phase transition.
Conversely, in the physical portion of parameter space, all fluctuations have positive $M^2>0$,
and are never parametrically light.

By inspecting Figs.~\ref{Fig:mass1},  \ref{Fig:mass2},  \ref{Fig:mass3},  and~\ref{Fig:mass4},
we notice that the mass of the lightest state in the spectrum, in units of the mass of the lightest tensor, becomes smaller
when the source increases towards its critical value.
Hence, the minimum value $M(c)$ of the mass of the lightest scalar is found in immediate proximity of the transition.
For each choice of $\Delta$, we compute the mass spectrum of bound states of the dual theory
precisely at the phase transition point, which is displayed in Fig.~\ref{Fig:summary}.
We find two interesting, unexpected features, which we highlight in closing this section.

First,  the mass of the lightest scalar state, 
has a minimum in proximity of $\Delta=5/2$, in which case
the mass of such state is approximately one third of the mass of the tensor. 
We show explicitly in the top panel of Fig.~\ref{Fig:phase} the mass spectrum computed in the probe approximation, which shows significant disagreement with the mass of the lightest scalar
for large values of $\phi_I$.
In this case, the lightest scalar particle has a substantial overlap with the dilaton~\cite{Elander:2020csd}. 
Unfortunately, this feature appears only inside the metastable (or tachyonic) region(s) of parameter space,
while the probe approximation captures well the mass of the lightest scalar in the the physical region, when $|\phi_I| \leq |\phi_I(c)|$.

\begin{table}[t]
\caption{Summary table of our results. For each $\Delta$, 
we report the critical value of the normalised source 
$\hat{\phi}_{J}({\rm c})$ at the phase transition, 
the critical value of the IR expansion parameter $\phi_I({\rm c})$,
and the value of the mass {$M({\rm c})$} of the lightest scalar state
at the transition---expressed in units of the mass of the lightest spin-2 state for the same choices of parameters.
For completeness, we also report the value of the two parameters 
linked to the
condensates in the dual field  theory: $\hat{\phi}_{V}(c)$ and $\hat{\chi}_5(c)$.
 \label{Fig:table}\\}
\centering
\begin{tabular}{|c|c|c|c||c|c|}
\hline\hline
$~~~\Delta~~~$ & $~~~\hat{\phi}_{J}({\rm c})~~~$ & $~~~\phi_I({\rm c})~~~$ & {$~~~M({\rm c})~~~$} &
$~~~\hat{\phi}_{V}({\rm c})~~~$ & $~~~\hat{\chi}_{5}({\rm c})~~~$ \\
\hline
$~~0.50~~$ & $2.373$ & $1.89$ & $0.571$  & $-0.063$ & $-0.176$ \\
$~~1.00~~$ & $1.149$ & $0.845$ & $0.553$ & $-0.208$ & $-0.328$ \\
$~~1.50~~$ & $0.990$ & $0.567$ & $0.512$ & $-0.369$ & $-0.368$ \\
$~~1.70~~$ & $1.010$ & $0.494$ & $0.482$ & $-0.471$ & $-0.377$ \\
$~~1.75~~$ & $1.022$ & $0.477$ & $0.473$ & $-0.504$ & $-0.380$ \\
$~~1.80~~$ & $1.038$ & $0.459$ & $0.464$ & $-0.539$ & $-0.382$ \\
$~~1.85~~$ & $1.059$ & $0.442$ & $0.455$ & $-0.579$ & $-0.384$ \\
$~~1.90~~$ & $1.084$ & $0.424$ & $0.446$ & $-0.625$ & $-0.386$ \\
$~~1.95~~$ & $1.115$ & $0.407$ & $0.436$ & $-0.677$ & $-0.388$ \\
$~~2.00~~$ & $1.153$ & $0.388$ & $0.427$ & $-0.736$ & $-0.390$ \\
$~~2.25~~$ & $1.567$ & $0.281$ & $0.385$ & $-1.276$ & $-0.401$ \\
$~~2.30~~$ & $1.756$ & $0.254$ & $0.379$ & $-1.496$ & $-0.403$ \\
$~~2.35~~$ & $2.045$ & $0.223$ & $0.374$ & $-1.821$ & $-0.405$ \\
$~~2.40~~$ & $2.554$ & $0.186$ & $0.370$ & $-2.371$ & $-0.408$ \\
$~~2.45~~$ & $3.748$ & $0.136$ & $0.367$ & $-3.618$ & $-0.410$ \\
$~~2.49~~$ & $8.750$ & $0.063$ & $0.368$ & $-8.691$ & $-0.412$ \\
$~~2.499~~$ & $27.82$ & $0.020$ & $0.369$ & $-27.80$ & $-0.413$ \\
$~~2.50~~$ & $-0.295$ & $0.107$ & $0.359$ & $0.439$ & $-0.411$ \\
$~~2.501~~$ & $147.3$ & $0.107$ & $0.359$ & $-147.2$ & $-0.411$ \\
$~~2.51~~$ & $14.79$ & $0.107$ & $0.362$ & $-14.69$ & $-0.411$ \\
$~~2.55~~$ & $3.001$ & $0.108$ & $0.372$ & $-2.899$ & $-0.411$ \\
$~~2.60~~$ & $1.528$ & $0.108$ & $0.384$ & $-1.426$ & $-0.411$ \\
$~~2.65~~$ & $1.038$ & $0.108$ & $0.397$ & $-0.935$ & $-0.411$ \\
$~~2.70~~$ & $0.793$ & $0.109$ & $0.409$ & $-0.689$ & $-0.411$ \\
$~~2.75~~$ & $0.647$ & $0.109$ & $0.421$ & $-0.542$ & $-0.411$ \\
$~~3.00~~$ & $0.358$ & $0.113$ & $0.481$ & $-0.246$ & $-0.411$ \\
$~~3.50~~$ & $0.225$ & $0.127$ & $0.586$ & $-0.068$ & $-0.411$ \\
$~~4.00~~$ & $0.189$ & $0.150$ & $0.616$ & $-0.102$ & $-0.410$ \\
$~~4.80~~$ & $0.136$ & $0.204$ & $0.620$ & $-0.179$ & $-0.410$ \\
\hline
\end{tabular}
\end{table}

Second, this summary plot displays a peculiar discontinuity at $\Delta=5/2$
(see the inset of Fig.~\ref{Fig:summary}). The numerical value for the mass of the lightest scalar at $\Delta=5/2$ aligns well with those obtained for $\Delta>5/2$,
and is the absolute minimum of the mass. However, the sequence of masses of the lightest scalar state one obtains for $\Delta<5/2$ converges to a slightly higher value.
We do not know why this second feature emerges, and highlight to the reader the intrinsic 
numerical difficulty of some of the analysis we carried out.
Yet, this discontinuity is a small effect, when compared to the much more significant 
first feature we found, namely that the mass of the lightest scalar is minimised for $\Delta=5/2$.

\begin{figure}[t]
\begin{center}
\includegraphics[width=17cm]{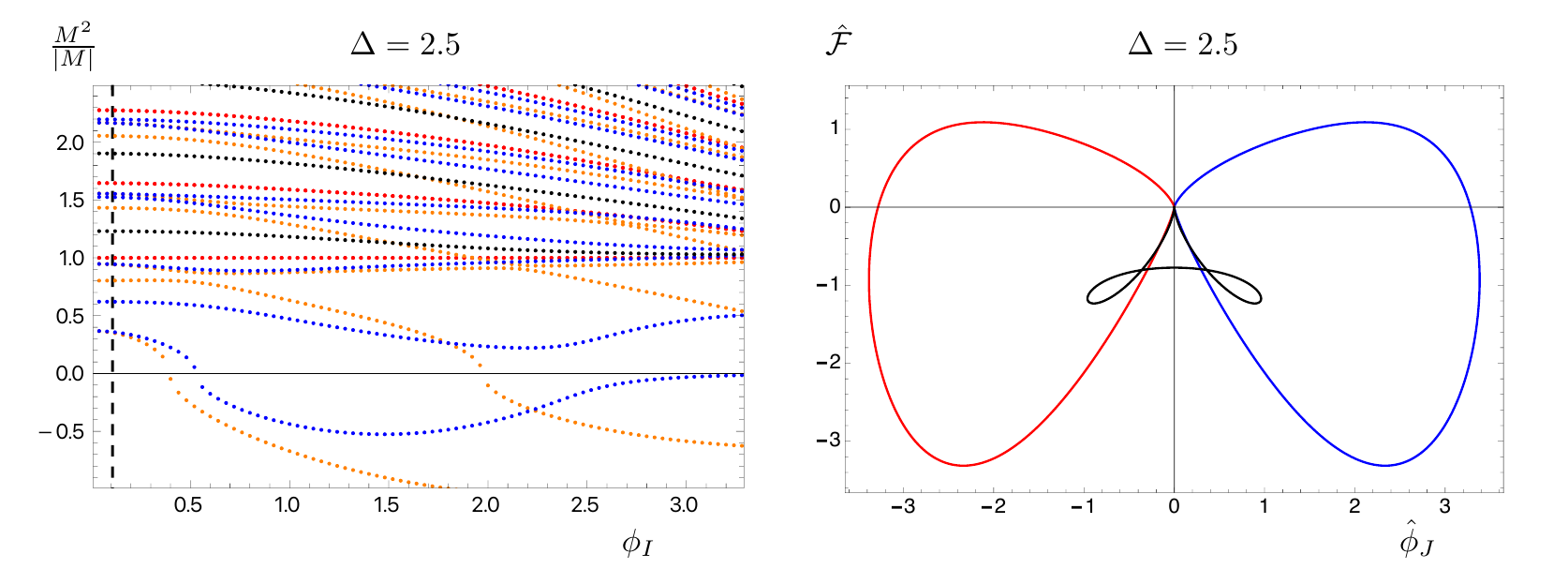}
\caption{
  The mass spectrum of fluctuations  for $\Delta=5/2$,
displaying the spin-0 (blue), spin-1 (black), and spin-2 (red) states, in units of the
mass of the lightest spin-2 state (left panel).
The spectrum also includes the masses of the scalars computed in the probe approximation (orange),
for comparison with the complete calculation (blue). The values of the IR and UV cutoffs in the calculations are respectively given by $\rho_1-\rho_o=10^{-9}$ and  $\rho_2-\rho_o=5$.
  The (normalised) free energy density $\hat{\cal F}$, as a function of the source $\hat{\phi}_{J}$,
  for $\Delta=5/2$ (right panel). The black curve denotes the confining solutions, while the red and blue solutions are singular 
domain-wall ones.
 The vertical, dashed line in the left panel indicates the position of the first-order phase transition 
  evident in the right panel.}
\label{Fig:phase}
\end{center}
\end{figure}

\begin{figure}[th]
\begin{center}
\includegraphics[width=12cm]{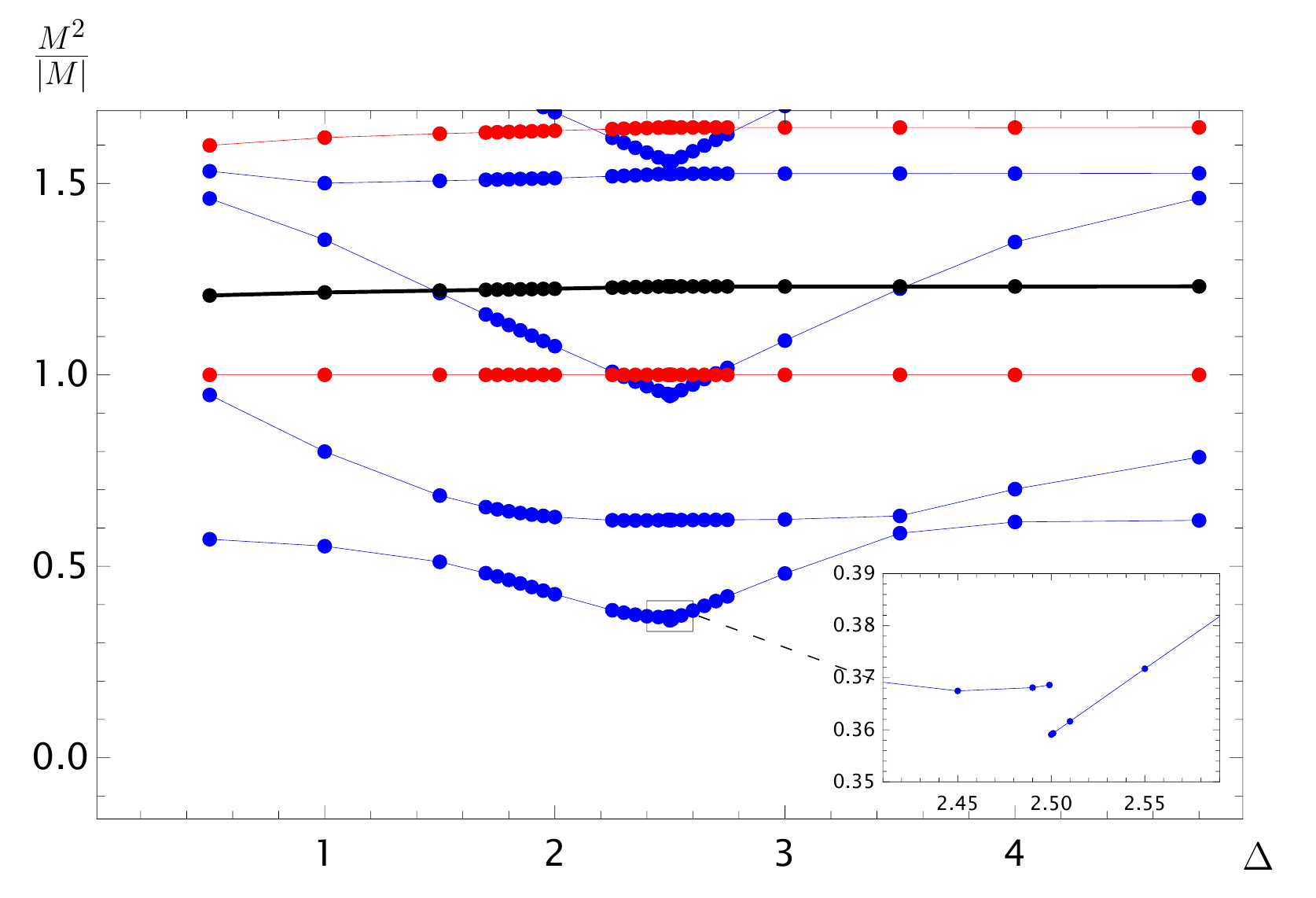}
\caption{The mass spectrum of fluctuations of the gravity theory (corresponding to the bound states of the dual field theory) evaluated at the critical $\phi_I(c)$ for each choice of $\Delta$. We display the spin-0 (blue),
spin-1 (black), and spin-2 (red) states. The values of the IR and UV cutoffs in the calculations are respectively given by $\rho_1-\rho_o=10^{-9}$ and  $\rho_2-\rho_o=5$. All masses are normalised to the mass of the lightest spin-2 state.
} 
\label{Fig:summary}
\end{center}
\end{figure}

\section{Outlook}
\label{Sec:outlook}

In a broad class of bottom-up holographic models,
in which the background geometry ends smoothly, so as to introduce a mass gap 
in the dual field theory in a way that mimics a confining theory, 
and furthermore in the presence of  an additional deformation (and 
the corresponding condensate), we found evidence
of the presence of a first-order phase transition.
The nature of the model allows one to treat $\Delta$, the dimensionality of the deformation/condensate,
as a free parameter, and study the physics as a function of $\Delta$.
All the fluctuations of the gravity backgrounds
 have positive mass squared, $M^2>0$, for all values  considered, $0<\Delta<5$,
in the physical portion of the parameter space, before 
the appearance of the phase transition.
Interestingly, in portions of the
parameter space well beyond the phase transition, 
the lightest scalar state in the theory becomes first an approximate dilaton,
its mass being paramaterically suppressed, and then a tachyon.
All of these results confirm and generalise 
the findings in the three  top-down models studied in Refs.~\cite{Elander:2020ial,Elander:2020fmv,Elander:2021wkc},
 as well as the 
expectations of Refs.~\cite{Gorbenko:2018ncu,Gorbenko:2018dtm,Pomarol:2019aae}.

Restricting attention to the physical region 
of parameter space, in which the confining solutions are 
stable, we found some evidence that the mass of the lightest state in the spectrum
becomes smaller in the limit  $\Delta\rightarrow 5/2$--- half of the dimension of the space-time of the CFT that the dual theory flows from in the UV---in line with the suggestion,  in Ref.~\cite{Kaplan:2009kr}, that such values of $\Delta$ are special.
However, the lightest mass is light, but not parametrically so, as predicted in Ref.~\cite{Pomarol:2019aae}.

This study suggests that the phase transition we uncovered in this class of models
 cannot be rendered arbitrarily weak,
nor of second order, and the lightest state cannot be a dilaton
with tunable, small mass.
The phenomena associated with 
confinement in the dual field theory, and its specific implementation---which we
borrowed from Witten's model~\cite{Witten:1998zw}---are 
playing an important role in yielding this conclusion.
These findings support the arguments according to which
inferring the properties of  long-distance dynamics on the basis solely 
of the short-distance characterisation of the theory (its fixed points, and the classification of their
deformations), is misleading, as the RG flow away from fixed points, ultimately leading to confinement, adds 
important non-perturbative effects.

It is premature to try to generalise these conclusions: the class of models 
analysed here does not cover all possible 
realisations of confining field theories, and possibly the limitation of all these studies descends from the fact
that confinement is described always in the same way, in terms of a shrinking circle in the internal geometry
of a higher-dimensional gravity theory.
For instance, top-down holographic models in which the analysis of the spectrum of fluctuations has been performed include Refs.~\cite{Elander:2018gte}, and~\cite{Elander:2017cle,Elander:2017hyr}, 
and the results (in particular, regarding the presence of light states) are model-dependent, as the mechanism introducing 
a mass gap is different from the mechanism of interest here. We remind the Reader that in Refs.~\cite{Elander:2020fmv, Elander:2021kxk} a similar model, but derived from maximal supergravity in seven dimensions, was studied. The main difference here is the fact that, in the bottom-up model considered in the present paper, $\Delta$ is a free parameter. The search for models that realise a weak phase transition, and in which a parametrically light dilaton can be realised, in the presence of a nearby instability in the parameter space of the theory, continues.

We conclude by anticipating that we will
further extend this six-dimensional model,
along lines similar  to those followed in Ref.~\cite{Elander:2021kxk},
in the complementary context  of composite Higgs models,
 in future publications~\cite{SO(5),vacuum,EWSB}.
In this different context, we will extend the sigma-model to incorporate an internal gauge symmetry.
We will study how to break the corresponding global symmetry in the field theory, 
how to (weakly) gauge a subgroup, and how to exploit vacuum (mis-)alignment 
to construct a composite Higgs model, with semi-realistic features---in particular
in reference to electroweak symmetry breaking.

\begin{acknowledgments}

We would like to thank Ed Bennett for assisting in  the preparation of
the data release.

The work of AF has been supported by the STFC 
Consolidated Grant No. ST/V507143/1.

The work of  MP has been supported in part by the STFC 
Consolidated Grants No. ST/P00055X/1 and No. ST/T000813/1.
  MP received funding from
the European Research Council (ERC) under the European
Union’s Horizon 2020 research and innovation program
under Grant Agreement No.~813942.

\vspace{1.0cm}

{\bf Research Data - } All data displayed in the figures of this manuscript can be downloaded from Ref.~\cite{ZEAP}. We refer the reader to Refs.~\cite{Bennett:2022klt,Athenodorou:2022ixd} and references therein for recent surveys on open science.

\vspace{1.0cm}

{\bf Open Access Statement - } For the purpose of open access, the authors have applied a Creative Commons 
Attribution (CC BY) licence  to any Author Accepted Manuscript version arising.

\end{acknowledgments}

\newpage
\appendix

\section{Sigma-model coupled to gravity}
\label{Sec:DAction}

For the purpose of fixing our conventions, we write here 
the action of the two-derivative sigma-model in $D$ dimensions, consisting of $n$ scalars $\Phi^a$, with $a=1,\,\cdots,\,n$, coupled to gravity:
\beqs
\label{eq:sigmamodelaction}
	\mathcal S &= \int \dd^D x \sqrt{-g} \, \bigg[ \frac{R}{4} -
	 \frac{1}{2} g^{MN}G_{ab} \partial_M \Phi^a \partial_N \Phi^b - {\mathcal V}(\Phi^a)\bigg] \,.
	  \eeqs
We denote by $M = 0,\cdots,\,3,\,5,\,\cdots,\,D$ the $D$-dimensional space-time indexes. The $D$-dimensional metric $g_{MN}$ has determinant $g$, and signature mostly $+$. The $D$-dimensional Ricci scalar is denoted by $R$. The sigma-model metric is $G_{ab}$, and its inverse is $G^{ab}$. The potential $\mathcal V$ is a function of the scalars $\Phi^a$. 
 
If we adopt the ansatz for the background solutions, that the  metric is of the DW form, and that the scalar fields only depend on the radial coordinate,
\beqs
\dd s_D^2 &=& \dd r^2 + e^{2{A}(r)} \dd x_{1,D-2}^2\,,\\
\Phi^a&=&\Phi^a(r)\,,
\eeqs
then the equations of motion are given by 
\beqs
\label{Eq:scalarEQ}
\partial_r^2\Phi^a + (D-1)\partial_r A \partial_r \Phi^a + {\mathcal G}^a_{\,\,\,\,bc} \partial_r \Phi^b \partial_r \Phi^c - G^{ab} \frac{\partial \mathcal V}{\partial \Phi^b}
&=& 0\,,\\
\label{Eq:Einstein1}
(D-1) \left( \partial_r A \right)^2 + \partial_r^2 A + \frac{4}{D-2} \mathcal V &=& 0\,,\\
\label{Eq:Einstein2}
(D-1)(D-2) \left( \partial_r A \right)^2 - 2 G_{ab} \partial_r \Phi^a \partial_r \Phi^b + 4 \mathcal V &=& 0\,,
\eeqs
where the sigma-model connection is
\beqs
\label{Eq:sigma}
{\cal G}^d_{\,\,\,\,ab}&\equiv& \frac{1}{2}G^{dc}\left(\frac{}{}\partial_aG_{cb}+\partial_bG_{ca}-\partial_cG_{ab}\right)\,.
\eeqs

Furthermore, if one finds a solution, $\mathcal W(\Phi^a)$, to the following partial differential equation:
\beqs
	\mathcal V &= &\frac{1}{2} G^{ab} \frac{\partial \mathcal W}
	{\partial \Phi^a}\frac{\partial \mathcal W}
	{\partial \Phi^b} - \frac{D-1}{D-2} \mathcal W^2 \,,
\eeqs
then any solution to the first-order equations
\beqs
\label{Eq:W1}
\partial_r A &=& -\frac{2}{D-2} \mathcal W\,,\\
\label{Eq:W2}
\partial_r \Phi^a &=& G^{ab}\frac{\partial \mathcal W}{\partial \Phi^b}\,,
\eeqs
is also a solution of the second-order classical Eqs.~(\ref{Eq:scalarEQ}), (\ref{Eq:Einstein1}), and~(\ref{Eq:Einstein2}).  

\subsection{Linearised equations for scalar  fluctuations}
\label{Sec:Afluctutations}

In general, the linearised equations of motion can be put in gauge-invariant form following the formalism in Refs.~\cite{Bianchi:2003ug,Berg:2005pd,Berg:2006xy,Elander:2009bm,Elander:2010wd} (where more details can be found).
For the scalars, the gauge-invariant fluctuations $\mathfrak{a}^{a}$ obey the following equation
\beqs
\label{eq:Ascalarflucs1}
\Big[ \mathcal{D}_{r}^2 + (D-1)\partial_{r}A\,\mathcal{D}_{r}
-e^{-2A}q^{2} \Big] \mathfrak{a}^{a} - \mathcal X^{a}_{\ c}
\mathfrak{a}^{c}&=&0\,.
\eeqs
Here, we used the notation that, given a field $X^a$, 
  the sigma-model covariant, and background-covariant, derivatives are defined
   by $D_b X^a = \partial_b X^a + \mathcal G^a{}_{bc} X^c$ and $\mathcal D_r X^a 
   = \partial_r X^a + \mathcal G^a{}_{bc} \partial_r \Phi^b X^c$, respectively, in terms of the sigma-model connection $\mathcal G^a{}_{bc}$ of Eq.~(\ref{Eq:sigma}). 
   The sigma-model Riemann tensor is then given by $\mathcal R^a{}_{bcd} = \partial_c \mathcal G^a{}_{bd} - \partial_d \mathcal G^a{}_{bc} + \mathcal G^a{}_{ce} \mathcal G^e{}_{bd} - \mathcal G^a{}_{de} \mathcal G^e{}_{bc}$.
   The matrix  $\mathcal X^{a}_{\ c}$ reads as follows: 
   \beqs
\label{eq:Ascalarflucs2}
\mathcal X^{a}_{\ c}&\equiv&
-\mathcal{R}^{a}_{\ bcd}\partial_{r} \Phi^{b}\partial_{r} \Phi^{d} 
+D_c \bigg(G^{ab}\frac{\partial \mathcal V}{\partial  \Phi^{b}}\bigg)\, 
+\frac{4}{(D-2)\partial_{r} A}
\bigg[\partial_{r} \Phi^{a}\frac{\partial \mathcal V}{\partial  \Phi^{c}}+G^{ab}\frac{\mathcal V}{\partial  \Phi^{b}}\partial_{r}\Phi^{d}G_{dc}\bigg]\nonumber\\
&&
+\frac{16 \mathcal V}{(D-2)^2(\partial_{r}A)^{2}}\partial_{r} \Phi^{a}\partial_{r} \Phi^{b}G_{bc}\, .
\eeqs

We add infinitely large, boundary-localised mass terms for all the fluctuations of $\Phi^a$, that hence satisfy Dirichlet boundary conditions.  The resulting boundary conditions for  the gauge-invariant variables $\mathfrak a^a$ are the following:
\be
	\label{eq:Ascalarflucs3}
	\partial_{r} \Phi^{c}
	\partial_{r} \Phi^{d}G_{db}{\cal D}_{\rho}\mathfrak{a}^{b}\bigg|_{r_{i}}
	= \left[ \frac{(D-2)\partial_{r}A}{2}e^{-2A}q^{2}\delta^{c}_{\ b}
	+\partial_{r} \Phi^{c}\bigg(\frac{4\mathcal V}{(D-2)\partial_{r}A}\partial_{r} \Phi^{d}G_{db} + \frac{\partial\mathcal V}{\partial\Phi^{b}} \bigg) \right]\mathfrak{a}^{b}\bigg|_{r_{i}}\, .
\ee

\section{Of superpotentials}
\label{Sec:super}

The associated superpotential, Eq.~(\ref{Eq:AW}), that is used for example in the context of holographic renormalisation when $\Delta>5/2$,
is not known in closed form. Yet, we can write it as a power expansion in $\phi$, and we find that:
\beqs
\overline{\mathcal W}_6
&=&-2-\frac{1}{2}(5-\Delta) {\phi}^2
-\frac{25 (2 \Delta -5)}{16 (4 \Delta -15)}{\phi}^4
-\frac{125 (2 \Delta -5) \left(4 \Delta ^2-15 \Delta +25\right)}{64 (4 \Delta -15)^2 (6
   \Delta -25)}{\phi}^6\\
   && -\frac{625 \left(128 \Delta ^5-480 \Delta ^4+800 \Delta ^3-4\times10^3 \Delta ^2+11250 \Delta
   -9375\right)}{1024 (4 \Delta -15)^3 \left(48 \Delta ^2-410 \Delta
   +875\right)}\, \phi^8\nonumber\\
      && -\frac{625}{4096 (25-6 \Delta )^2 (15-4 \Delta )^4 \left(16 \Delta
   ^2-142 \Delta +315\right)}\left(3072 \Delta ^8+25600 \Delta ^7-433600 \Delta ^6
  \frac{}{} \right.\nonumber\\
   &&\left.\frac{}{}
   +1292\times10^3 \Delta
   ^5+1065\times10^3\Delta ^4-10862500 \Delta ^3+18531250 \Delta ^2-12734375 \Delta
   +3515625\right) \,\phi^{10}\nonumber\\
         &&-\frac{3125 }{32768 (25-6
   \Delta )^2 (4 \Delta -15)^5 \left(192 \Delta ^3-2584 \Delta ^2+11590 \Delta
   -17325\right)}
 \left(\frac{}{}24576 \Delta ^{10}+1320960 \Delta ^9 \frac{}{} \right.\nonumber\\
   &&\left.\frac{}{}
   -8153600 \Delta ^8-45968\times10^3\Delta^7 
   +473\times10^6 \Delta ^6-12998\times10^5 \Delta ^5+100425\times10^4 \Delta ^4
    \frac{}{} \right.\nonumber\\
   &&\left.\frac{}{}
   +133625\times10^4 \Delta
   ^3-25625\times10^5 \Delta ^2+10^9 \Delta +87890625\frac{}{}\right)\, \phi^{12}   \nonumber\\
            &&
            -\frac{15625}{131072 (35-8 \Delta )^2 (15-4 \Delta )^6 (6 \Delta -25)^3 \left(336 \Delta
   ^3-4612 \Delta ^2+21100 \Delta -32175\right)}
    \left(\frac{}{}4718592 \Delta ^{14}
    \frac{}{} \right.\nonumber\\
   &&\left.\frac{}{}
+714670080 \Delta ^{13}-2224128\times10^3 \Delta
   ^{12}-140363776\times10^3 \Delta ^{11}+155181568\times10^4 \Delta ^{10}-54067432\times10^5 \Delta
   ^9
   \frac{}{} \right.\nonumber\\
   &&\left.\frac{}{}
-6193644\times10^6 \Delta ^8+107537295\times10^6 \Delta ^7-3559741875\times10^5 \Delta
   ^6+53236421875\times10^4 \Delta ^5
   \frac{}{} \right.\nonumber\\
   &&\left.\frac{}{}-2431853125\times10^5 \Delta ^4
-318709765625\times10^3 \Delta
   ^3+450370605468750 \Delta ^2
   \frac{}{} \right.\nonumber\\
   &&\left.\frac{}{}-162176513671875 \Delta +823974609375\frac{}{}\right)\, \phi^{14}
   \nonumber\\
               &&
                -\frac{78125}{4194304 (35-8 \Delta )^2 (4 \Delta -15)^7 (6
   \Delta -25)^3 \left(5376 \Delta ^4-98992 \Delta ^3+683500 \Delta ^2-2097300 \Delta
   +2413125\right)}\times
   \frac{}{} \nonumber\\
   &&\frac{}{} \times
    \left(\frac{}{}150994944 \Delta ^{16}
   +57881395200 \Delta ^{15}+1092472012800 \Delta
   ^{14}
        \frac{}{} \right.\nonumber\\
   &&\left.\frac{}{}
   -27151204352\times10^3 \Delta ^{13}+8460660736\times10^4 \Delta ^{12}
   +13597392384\times10^5 \Delta
   ^{11}
        \frac{}{} \right.\nonumber\\
   &&\left.\frac{}{}
   -13484257056\times10^6 \Delta ^{10}+4795751472\times10^7 \Delta ^9-398284926\times10^8
   \Delta ^8
     \frac{}{} \right.\nonumber\\
   &&\left.\frac{}{}
   -254111197\times10^9 \Delta ^7+9657946325\times10^8 \Delta
   ^6-14713569328125\times10^5 \Delta ^5
        \frac{}{} \right.\nonumber\\
   &&\left.\frac{}{}
   +88646595703125\times10^4 \Delta ^4
   +26974482421875\times10^4
   \Delta ^3-617424755859375\times10^3 \Delta ^2
        \frac{}{} \right.\nonumber\\
   &&\left.\frac{}{}
   +224028259277343750 \Delta
   +61798095703125\frac{}{}\right)\, \phi^{16}   \nonumber\\
   &&
   -{390625}\left[\frac{}{}16777216
   (25-6 \Delta )^4 (15-4 \Delta )^8 \left(16 \Delta ^2-142 \Delta +315\right)^2 \times
    \frac{}{} \right.\nonumber\\
   &&\left.\frac{}{}\times
   \left(48384 \Delta ^4-901680 \Delta ^3+6301100 \Delta ^2-19569500 \Delta
   +22790625\right)\right]^{-1}
   \times
   \frac{}{} \nonumber\\
   &&\frac{}{} \times
 \left(7247757312 \Delta ^{20}+6263874256896 \Delta ^{19}+303911927808\times10^3
   \Delta ^{18}
        \frac{}{} \right.\nonumber\\
   &&\left.\frac{}{}
   -5902748929228800 \Delta ^{17}-2758035111936\times10^4 \Delta
   ^{16}+13174340632576\times10^5 \Delta ^{15}
        \frac{}{} \right.\nonumber\\
   &&\left.\frac{}{}
   -115787186756608\times10^5 \Delta
   ^{14}+31365013166976\times10^6 \Delta ^{13}+17161524832384\times10^7 \Delta
   ^{12}
        \frac{}{} \right.\nonumber\\
   &&\left.\frac{}{}
   -18328634352128\times10^8 \Delta ^{11}+726376592369\times10^{10} \Delta
   ^{10}-140457265478675\times10^8 \Delta ^9
        \frac{}{} \right.\nonumber\\
   &&\left.\frac{}{}
   +3591879533675\times10^9 \Delta
   ^8+4946038879446875\times10^7 \Delta ^7-126333638659140625\times10^6 \Delta^6
        \frac{}{} \right.\nonumber\\
   &&\left.\frac{}{}
   +1469488973640625\times10^8 \Delta ^5-7281017740966796875\times10^4 \Delta
   ^4-14333258841552734375\times10^3 \Delta ^3
        \frac{}{} \right.\nonumber\\
   &&\left.\frac{}{}
   +322610068359375\times10^8 \Delta
   ^2-10350940532684326171875 \Delta +115871429443359375\right)\, \phi^{18}   \nonumber\\
   &&
   -{390625}
   \left[\frac{}{}134217728 (25-6 \Delta )^4 (9-2 \Delta )^2 (4 \Delta -15)^9 (8 \Delta -35)^3\times
           \frac{}{} \right.\nonumber\\
   &&\left.\frac{}{}\times
   \left(193536 \Delta ^5-4526016 \Delta ^4+42336320 \Delta ^3-197998900 \Delta
   ^2+462983\times10^3 \Delta -433021875\right)\frac{}{}\right]^{-1}
      \times
   \frac{}{} \nonumber\\
   &&\frac{}{} \times
    \left(\frac{}{}
    463856467968 \Delta ^{23}+870020787732480 \Delta
   ^{22}+93468471475568640 \Delta ^{21}
           \frac{}{} \right.\nonumber\\
   &&\left.\frac{}{}
   -537522325998796800 \Delta
   ^{20}-4220626598363136\times10^4 \Delta ^{19}+70461689518882816\times10^4 \Delta
   ^{18}
           \frac{}{} \right.\nonumber\\
   &&\left.\frac{}{}
   -24426057346514944\times10^5 \Delta ^{17}-38568233605083136\times10^6 \Delta
   ^{16}           \frac{}{} \right.\nonumber\\
   &&\left.\frac{}{}
   +51863822701075456\times10^7 \Delta ^{15}
   -27603786074733376\times10^8 \Delta
   ^{14}
              \frac{}{} \right.\nonumber\\
   &&\left.\frac{}{}
   +5554033898884608\times10^9 \Delta ^{13}+1756207389007328\times10^{10} \Delta
   ^{12}
           \frac{}{} \right.\nonumber\\
   &&\left.\frac{}{}
   -1601187067059638\times10^{11} \Delta ^{11}+53057037376123675\times10^{10} \Delta
   ^{10}
              \frac{}{} \right.\nonumber\\
   &&\left.\frac{}{}
   -922419869075040625\times10^9 \Delta ^9
   +5120596044251203125\times10^8 \Delta
   ^8
              \frac{}{} \right.\nonumber\\
   &&\left.\frac{}{}
   +138753717\times10^3403125\times10^{10} \Delta ^7-371419078250947265625\times10^7 \Delta
   ^6
           \frac{}{} \right.\nonumber\\
   &&\left.\frac{}{}
   +4075653348253369140625\times10^6 \Delta ^5-20041289228924560546875\times10^5 \Delta
   ^4
              \frac{}{} \right.\nonumber\\
   &&\left.\frac{}{}
   -9840867372802734375\times10^7 \Delta ^3   
   +54259196491241455078125\times10^4 \Delta
   ^2
              \frac{}{} \right.\nonumber\\
   &&\left.\frac{}{}
   -169888076331138610839843750 \Delta -60832500457763671875\frac{}{}\right)
   \, \phi^{20}
   \nonumber\\
+\cdots\,.\nonumber
\eeqs
We show explicitly terms up to ${\cal O}(\phi^{20})$ in order to allow the reader to
inspect the presence of a pathology in the expansion: for special choices such
as $\Delta=\frac{15}{4}, \frac{25}{6}, \frac{35}{8}, \frac{9}{2}, \frac{55}{12}, \frac{65}{14}, 
\frac{75}{16},\frac{85}{18}, \frac{19}{4}, \cdots$, some coefficients  diverge.

\section{Asymptotic expansion of the background solutions}
\label{Sec:UVsolutions}

In the main body of the paper, we showed the IR expansion for the confining solutions for any value of $\Delta$,
in Eqs.~(\ref{Eq:IR1}), (\ref{Eq:IR2}), and~(\ref{Eq:IR3}). For the singular domain-wall solutions,
we provided the IR expansions in Eqs.~(\ref{Eq:IR4}) and~(\ref{Eq:IR5}).
Contrary to the IR expansions, the functional form of the UV expansions depends explicitly on $\Delta$.
In the main body of the paper, we reported the expansion for $\Delta=3$. 
In this Appendix, we provide a few additional representative examples of the UV expansions,
expressed as power series in $z=e^{-\r}$. For simplicity, we set $A_U=0=\chi_U$.

For $\Delta=1/2$, and hence $\Delta_J=1/2$, $\Delta_V=9/2$, we have

\beqs
    \phi &=& z^{1/2} \phi_J + z^{9/2} \phi _V+\frac{3}{5} z^{11/2} \phi _J^2 \phi
   _V+\frac{35}{192} z^{13/2} \phi _J^4 \phi _V + o(z^7) \,, \\
    \chi &=& -\frac{\log (z)}{3}-\frac{z \phi _J^2}{24}+z^5 \left(\text{$\chi
   $5}-\frac{3 \phi _J \phi _V}{100}\right)+z^6 \left(\frac{25 \text{$\chi
   $5} \phi _J^2}{48}-\frac{11}{720} \phi _J^3 \phi _V\right) \nonumber \\ && +z^7
   \left(\frac{125 \text{$\chi $5} \phi _J^4}{896}-\frac{65 \phi _J^5 \phi
   _V}{16128}\right) + o(z^7) \,, \\
    A &=& -\frac{4 \log (z)}{3}-\frac{z \phi _J^2}{6}+z^5 \left(\frac{\text{$\chi
   $5}}{4}-\frac{3 \phi _J \phi _V}{25}\right)+z^6 \left(\frac{25
   \text{$\chi $5} \phi _J^2}{192}-\frac{11}{180} \phi _J^3 \phi
   _V\right) \nonumber \\ && +z^7 \left(\frac{125 \text{$\chi $5} \phi _J^4}{3584}-\frac{65
   \phi _J^5 \phi _V}{4032}\right) + o(z^7) \,.
\eeqs


For $\Delta=1$, and hence $\Delta_J=1$, $\Delta_V=4$, we have
\beqs
\phi&=&
z \phi _{J}
+z^4 \phi_{V}
+\frac{21}{40} z^6 \phi _{J}^2 \phi _{V}
         \,+\,{ o}(z^7)
\,,\\
\chi&=&-\frac{\log (z)}{3}
-\frac{1}{24} z^2 \phi _{J}^2
+\frac{1}{75} z^5 \left(75 \chi _5-4 \phi _{J} \phi_{V}\right)
-\frac{1}{280} z^7 \phi _{J}^2 \left(6 \phi _{J} \phi _{V}
-125  \chi _5\right)
         \,+\,{ o}(z^7)
\,,\\
A&=&-\frac{4 \log (z)}{3}
   -\frac{1}{6} z^2 \phi _{J}^2
   +\frac{1}{300} z^5 \left(75 \chi _5-64 \phi _{J} \phi_{V}\right)
   +\frac{z^7 \left(125 \chi _5 \phi _{J}^2-96 \phi _{J}^3 \phi_{V}\right)}{1120}
         \,+\,{ o}(z^7)
\,.
\eeqs


For $\Delta=3/2$, and hence $\Delta_J=3/2$, $\Delta_V=7/2$, we have
\beqs
\phi&=& z^{3/2} \phi_{J}
+z^{7/2} \phi _{V}
+\frac{2}{5} z^{13/2} \phi _{J}^2 \phi _{V}
         \,+\,{ o}(z^7)
\,,\\
\chi&=&-\frac{\log (z)}{3}
-\frac{1}{24} z^3 \phi _{J}^2
+\frac{1}{100} z^5 \left(100 \chi _5-7 \phi _{J} \phi_{V}\right)
-\frac{1}{24} z^7 \phi_{V}^2
         \,+\,{ o}(z^7)
\,,\\
A&=&-\frac{4 \log (z)}{3}
   -\frac{1}{6} z^3 \phi _{J}^2
   +\frac{1}{100} z^5 \left(25 \chi _5-28 \phi _{J} \phi_{V}\right)
   -\frac{1}{6} z^7 \phi_{V}^2
         \,+\,{ o}(z^7)
\,.
\eeqs


For $\Delta=17/10$, and hence $\Delta_J=17/10$, $\Delta_V=33/10$, we have
\beqs
\phi&=&
z^{17/10} \phi _{J} +z^{33/10} \phi _{V}+\frac{42}{125} z^{67/10} \phi _{J} ^2 \phi _{V} 
         \,+\,{ o}(z^7)
\,,\\
\chi&=&-\frac{\log (z)}{3}-\frac{1}{24} z^{17/5} \phi _{J} ^2+z^5 \left(\chi_5-\frac{187 \phi _{J}  \phi _{V} }{2500}\right) -\frac{1}{24} z^{33/5} \phi _{V} ^2
         \,+\,{ o}(z^7)
\,,\\
A&=&-\frac{4 \log (z)}{3}-\frac{1}{6} z^{17/5} \phi _{J} ^2+z^5 \left(\frac{\chi_5}{4}-\frac{187 \phi _{J}  \phi _{V} }{625}\right) -\frac{1}{6} z^{33/5} \phi _{V} ^2
         \,+\,{ o}(z^7)
\,.
\eeqs


For $\Delta=7/4$, and hence $\Delta_J=7/4$, $\Delta_V=13/4$, we have
\beqs
\phi&=&
z^{7/4} \phi _{J}+z^{13/4}\phi _{V}+\frac{51}{160} z^{27/4} \phi _{J} ^2 \phi _{V} 
         \,+\,{ o}(z^7)
\,,\\
\chi&=&-\frac{\log (z)}{3}-\frac{1}{24} z^{7/2} \phi _{J} ^2+z^5 \left(\chi_5-\frac{91 \phi _{J}  \phi _{V} }{1200}\right) {-\frac{1}{24} z^{13/2} \phi
   _V^2}
         \,+\,{ o}(z^7)
\,,\\
A&=&-\frac{4 \log (z)}{3}-\frac{1}{6} z^{7/2} \phi _{J} ^2+z^5 \left(\frac{\chi_5}{4}-\frac{91 \phi _{J}  \phi _{V} }{300}\right) {-\frac{1}{6} z^{13/2} \phi _V^2}
         \,+\,{ o}(z^7)
\,.
\eeqs


For $\Delta=9/5$, and hence $\Delta_J=9/5$, $\Delta_V=16/5$, we have
\beqs
\phi&=&
z^{9/5} \phi _{J}+z^{16/5}\phi _{V}+\frac{301 z^{34/5} \phi _{J} ^2 \phi _{V} }{1000}
         \,+\,{ o}(z^7)
\,,\\
\chi&=&-\frac{\log (z)}{3}-\frac{1}{24} z^{18/5} \phi _{J} ^2+z^5 \left(\chi_5-\frac{48 \phi _{J}  \phi _{V} }{625}\right) -\frac{1}{24} z^{32/5} \phi _{V} ^2
         \,+\,{ o}(z^7)
\,,\\
A&=&-\frac{4 \log (z)}{3}-\frac{1}{6} z^{18/5} \phi _{J} ^2+z^5 \left(\frac{\chi_5}{4}-\frac{192 \phi _{J}  \phi _{V} }{625}\right)-\frac{1}{6} z^{32/5} \phi _{V} ^2
         \,+\,{ o}(z^7)
\,.
\eeqs


For $\Delta=37/20$, and hence $\Delta_J=37/20$, $\Delta_V=63/20$, we have
\beqs
\phi&=&
z^{37/20} \phi _{J}+z^{63/20}\phi _{V}+\frac{1131 z^{137/20} \phi _{J} ^2 \phi _{V} }{4000}
         \,+\,{ o}(z^7)
\,,\\
\chi&=&-\frac{\log (z)}{3}-\frac{1}{24} z^{37/10} \phi _{J} ^2+z^5 \left(\chi_5-\frac{777 \phi _{J}  \phi _{V} }{10000}\right)-\frac{1}{24} z^{63/10} \phi _{V} ^2
         \,+\,{ o}(z^7)
\,,\\
A&=&-\frac{4 \log (z)}{3}-\frac{1}{6} z^{37/10} \phi _{J} ^2+z^5 \left(\frac{\chi_5}{4}-\frac{777 \phi _{J}  \phi _{V} }{2500}\right)-\frac{1}{6} z^{63/10} \phi _{V} ^2
         \,+\,{ o}(z^7)
\,.
\eeqs


For $\Delta=19/10$, and hence $\Delta_J=19/10$, $\Delta_V=31/10$, we have
\beqs
\phi&=&
z^{19/10} \phi _{J}+ { z^{31/10}\phi _{V} + \frac{33}{125} z^{69/10} \phi _{J} ^2 \phi _{V} }
         \,+\,{ o}(z^7)
\,,\\
\chi&=&-\frac{\log (z)}{3}-\frac{1}{24} z^{19/5} \phi _{J} ^2+z^5 \left(\chi_5-\frac{589 \phi _{J}  \phi _{V} }{7500}\right)-\frac{1}{24} z^{31/5} \phi _{V} ^2
         \,+\,{ o}(z^7)
\,,\\
A&=&-\frac{4 \log (z)}{3}-\frac{1}{6} z^{19/5} \phi _{J} ^2+z^5 \left(\frac{\chi_5}{4}-\frac{589 \phi _{J}  \phi _{V} }{1875}\right)-\frac{1}{6} z^{31/5} \phi _{V} ^2
         \,+\,{ o}(z^7)
\,.
\eeqs


For $\Delta=39/20$, and hence $\Delta_J=39/20$, $\Delta_V=61/20$, we have
\beqs
\phi&=&
z^{39/20} \phi _{J} +z^{61/20} \phi _{V}+\frac{979 z^{139/20} \phi _{J} ^2 \phi _{V} }{4000}+\,{ o}(z^7)
\,,\\
\chi&=&-\frac{\log (z)}{3}-\frac{1}{24} z^{39/10} \phi _{J} ^2+z^5 \left(\chi_5-\frac{793 \phi _{J}  \phi _{V} }{10000}\right)-\frac{1}{24} z^{61/10} \phi _{V} ^2
         \,+\,{ o}(z^7)
\,,\\
A&=&-\frac{4 \log (z)}{3}-\frac{1}{6} z^{39/10} \phi _{J} ^2+z^5 \left(\frac{\chi_5}{4}-\frac{793 \phi _{J}  \phi _{V} }{2500}\right)-\frac{1}{6} z^{61/10} \phi _{V} ^2
         \,+\,{ o}(z^7)
\,.
\eeqs


For $\Delta=2$, and hence $\Delta_J=2$, $\Delta_V=3$, we have
\beqs
\phi&=&
z^2 \phi _{J}+z^3 \phi
   _{\Delta _V}+\frac{9}{40} z^7 \phi _{J}^2 \phi _{V}
      \,+\,{ o}(z^7)
\,,\\
\chi&=&-\frac{\log (z)}{3}
-\frac{1}{24} z^4 \phi _{J}^2
+\frac{1}{25} z^5 \left(25 \chi _5-2 \phi _{J} \phi_{V}\right)
-\frac{1}{24} z^6 \phi_{V}^2
      \,+\,{ o}(z^7)
\,,\\
A&=&-\frac{4 \log (z)}{3}
-\frac{1}{6} z^4 \phi _{J}^2
+\frac{1}{100} z^5 \left(25 \chi _5-32 \phi _{J} \phi_{V}\right)
-\frac{1}{6} z^6 \phi_{V}^2
   \,+\,{ o}(z^7)
\,.
\eeqs

For $\Delta=9/4$, and hence $\Delta_J=9/4$, $\Delta_V=11/4$, we have
\beqs
\phi&=&
z^{9/4} \phi _{J}
+z^{11/4} \phi _{V}
   \,+\,{ o}(z^7)
\,,\\
\chi&=&
-\frac{\log (z)}{3}
-\frac{1}{24} z^{9/2} \phi _{J}^2
+\frac{1}{400} z^5 \left(400 \chi _5-33 \phi _{J} \phi_{V}\right)
-\frac{1}{24} z^{11/2} \phi_V^2
   \,+\,{ o}(z^7)
\,,\\
A&=&
-\frac{4 \log (z)}{3}
-\frac{1}{6} z^{9/2} \phi _{J}^2
+\frac{1}{100} z^5 \left(25 \chi _5-33 \phi _{J} \phi_{V}\right)
-\frac{1}{6} z^{11/2} \phi_{V}^2
   \,+\,{ o}(z^7)
\,.
\eeqs

For $\Delta=23/10$, and hence $\Delta_J=23/10$, $\Delta_V=27/10$, we have
\beqs
\phi&=&
z^{23/10} \phi _{J}+z^{27/10} \phi _{V}
   \,+\,{ o}(z^7)
\,,\\
\chi&=&
-\frac{\log (z)}{3}
-\frac{1}{24} z^{23/5} \phi _{J}^2
+\frac{z^5 \left(2500 \chi _5-207 \phi _{J} \phi_{V}\right)}{2500}
-\frac{1}{24} z^{27/5} \phi_{V}^2
   \,+\,{ o}(z^7)
\,,\\
A&=&
-\frac{4 \log (z)}{3}
-\frac{1}{6} z^{23/5} \phi _{J}^2
+\frac{z^5 \left(625 \chi _5-828 \phi _{J} \phi_{V}\right)}{2500}
-\frac{1}{6}   z^{27/5} \phi _{V}^2
   \,+\,{ o}(z^7)
\,.
\eeqs

{For $\Delta=47/20$, and hence $\Delta_J=47/20$, $\Delta_V=53/20$,} we have
\beqs
\phi&=&
z^{47/20} \phi _{J}+z^{53/20} \phi _{V}
   \,+\,{ o}(z^7)
\,,\\
\chi&=&
   -\frac{\log (z)}{3}
-\frac{1}{24} z^{47/10} \phi _{J}^2
+\frac{z^5 \left(30000 \chi _5-2491 \phi _{J} \phi_{V}\right)}{30000}
   -\frac{1}{24} z^{53/10}  \phi _{V}^2
   \,+\,{ o}(z^7)
\,,\\
A&=&
   -\frac{4 \log (z)}{3}
-\frac{1}{6} z^{47/10} \phi _{J}^2
+\frac{z^5 \left(1875 \chi _5-2491 \phi _{J} \phi_{V}\right)}{7500}
   -\frac{1}{6} z^{53/10} \phi_{V}^2
   \,+\,{ o}(z^7)
\,.
\eeqs

For $\Delta=12/5$, and hence $\Delta_J=12/5$, $\Delta_V=13/5$, we have
\beqs
\phi&=&
z^{12/5} \phi _{J}+z^{13/5} \phi _{V}
   \,+\,{ o}(z^7)
\,,\\
\chi&=&
 -\frac{\log (z)}{3}
 -\frac{1}{24} z^{24/5} \phi _{J}^2
+\frac{1}{625} z^5 \left(625 \chi _5-52 \phi _{J} \phi_{V}\right)
   -\frac{1}{24} z^{26/5} \phi_{V}^2
   \,+\,{ o}(z^7)
\,,\\
A&=&
  -\frac{4 \log (z)}{3}
     -\frac{1}{6} z^{24/5} \phi _{J}^2
+\frac{z^5 \left(625 \chi _5-832 \phi _{J} \phi_{V}\right)}{2500}
   -\frac{1}{6} z^{26/5} \phi_{V}^2
   \,+\,{ o}(z^7)
\,.
\eeqs

For $\Delta=49/20$, and hence $\Delta_J=49/20$, $\Delta_V=51/20$, we have
\beqs
\phi&=&
z^{49/20} \phi _{J}+z^{51/20} \phi _{V}
   \,+\,{ o}(z^7)
\,,\\
\chi&=&
   -\frac{\log (z)}{3}
     -\frac{1}{24} z^{49/10} \phi _{J}^2
+\frac{z^5 \left(10000 \chi _5-833 \phi _{J} \phi_{V}\right)}{10000}
   -\frac{1}{24} z^{51/10}\phi _{V}^2
   \,+\,{ o}(z^7)
\,,\\
A&=&
   -\frac{4 \log (z)}{3}
    -\frac{1}{6} z^{49/10} \phi _{J}^2
+\frac{z^5 \left(625 \chi _5-833 \phi _{J} \phi_{V}\right)}{2500}
   -\frac{1}{6} z^{51/10} \phi_{V}^2
   \,+\,{ o}(z^7)
\,.
\eeqs

{
For $\Delta=249/100$, and hence $\Delta_J=249/100$, $\Delta_V=251/100$, we have
\beqs
    \phi&=& z^{249/100} \phi _J+z^{251/100} \phi _V+ o(z^7) \,, \\
    \chi&=& -\frac{\log (z)}{3}-\frac{1}{24} z^{249/50} \phi _J^2+z^5
   \left(\text{$\chi $5}-\frac{20833 \phi _J \phi
   _V}{250000}\right)-\frac{1}{24} z^{251/50} \phi
   _V^2+ o(z^7) \,, \\
    A&=& -\frac{4 \log (z)}{3}-\frac{1}{6} z^{249/50} \phi _J^2+z^5
   \left(\frac{\text{$\chi $5}}{4}-\frac{20833 \phi _J \phi
   _V}{62500}\right)-\frac{1}{6} z^{251/50} \phi
   _V^2+ o(z^7) \,.
\eeqs

For $\Delta=2499/1000$, and hence $\Delta_J=2499/1000$, $\Delta_V=2501/1000$, we have
\beqs
    \phi&=& z^{2499/1000} \phi _J+z^{2501/1000} \phi _V+ o(z^7) \,, \\
    \chi&=& -\frac{\log (z)}{3}-\frac{1}{24} z^{2499/500} \phi _J^2+z^5
   \left(\text{$\chi $5}-\frac{2083333 \phi _J \phi
   _V}{25000000}\right)-\frac{1}{24} z^{2501/500} \phi
   _V^2+ o(z^7) \,, \\
    A&=& -\frac{4 \log (z)}{3}-\frac{1}{6} z^{2499/500} \phi _J^2+z^5
   \left(\frac{\text{$\chi $5}}{4}-\frac{2083333 \phi _J \phi
   _V}{6250000}\right)-\frac{1}{6} z^{2501/500} \phi
   _V^2+ o(z^7) \,.
\eeqs
}

For $\Delta=5/2$, and hence $\Delta_J=\Delta_V=5/2$, we have
\beqs
\phi&=&
\phi _{V} z^{5/2} 
+ \phi_{J} z^{5/2}
   \log (z)
+
\frac{1}{200}  \left(50 \phi_{V}^2 \phi_{J}-25 \phi_{V}
   \phi_{J}^2+8 \phi_{J}^3\right) z^{15/2}+\nonumber\\
   &&
   +\frac{1}{8} \left(4
   \phi_{V} \phi_{J}^2-\phi_{J}^3\right) z^{15/2} \log (z)
   +\frac{1}{4}  \phi_{J}^3 z^{15/2} \log ^2(z)
   \,+\,{ o}(z^8)
\,,\\
\chi&=&
   -\frac{\log (z)}{3}
+\left(\chi_5 +\frac{1}{600}(-25\phi_{V}^2-2\phi_{J}^2) \right)z^5
+\nonumber\\
&&
-\frac{1}{12}  \phi_{V} \phi_{J} z^5 \log
   (z)-\frac{1}{24}  \phi_{J}^2  z^5 \log ^2(z)
   \,+\,{ o}(z^8)
\,,\\
A&=&
   -\frac{4 \log (z)}{3}
+\frac{1}{300}  \left(75 \chi_5-50 \phi_{V}^2-4 \phi_{J}^2
   \right) z^5-\frac{1}{3}  \phi_{V} \phi_{J}  z^5 \log (z)-\frac{1}{6}
 \phi_{J}^2   z^5  \log ^2(z) \,+\,{ o}(z^8)
\,.
\eeqs

For $\Delta=2501/1000$, and hence $\Delta_J=2499/1000$, $\Delta_V=2501/1000$, we have
\beqs
    \phi&=& z^{2499/1000} \phi _J+z^{2501/1000} \phi _V+ o(z^7) \,, \\
    \chi&=& -\frac{\log (z)}{3}-\frac{1}{24} z^{2499/500} \phi _J^2+z^5
   \left(\text{$\chi $5}-\frac{2083333 \phi _J \phi
   _V}{25000000}\right)-\frac{1}{24} z^{2501/500} \phi
   _V^2+ o(z^7) \,, \\
    A&=& -\frac{4 \log (z)}{3}-\frac{1}{6} z^{2499/500} \phi _J^2+z^5
   \left(\frac{\text{$\chi $5}}{4}-\frac{2083333 \phi _J \phi
   _V}{6250000}\right)-\frac{1}{6} z^{2501/500} \phi
   _V^2+ o(z^7) \,.
\eeqs

For $\Delta=251/100$, and hence $\Delta_J=249/100$, $\Delta_V=251/100$, we have
\beqs
    \phi&=& z^{249/100} \phi _J+z^{251/100} \phi _V+ o(z^7) \,, \\
    \chi&=& -\frac{\log (z)}{3}-\frac{1}{24} z^{249/50} \phi _J^2+z^5
   \left(\text{$\chi $5}-\frac{20833 \phi _J \phi
   _V}{250000}\right)-\frac{1}{24} z^{251/50} \phi
   _V^2+ o(z^7) \,, \\
    A&=& -\frac{4 \log (z)}{3}-\frac{1}{6} z^{249/50} \phi _J^2+z^5
   \left(\frac{\text{$\chi $5}}{4}-\frac{20833 \phi _J \phi
   _V}{62500}\right)-\frac{1}{6} z^{251/50} \phi
   _V^2+ o(z^7) \,.
\eeqs

For $\Delta=51/20$, and hence $\Delta_J=49/20$, $\Delta_V=51/20$, we have
\beqs
\phi&=&
z^{49/20} \phi _{J}+z^{51/20} \phi _{V}
   \,+\,{ o}(z^7)
\,,\\
\chi&=&
   -\frac{\log (z)}{3}
      -\frac{1}{24} z^{49/10} \phi _{J}^2
+\frac{z^5 \left(10000 \chi _5-833 \phi _{J} \phi_{V}\right)}{10000}
   -\frac{1}{24} z^{51/10} \phi _{V}^2
   \,+\,{ o}(z^7)
\,,\\
A&=&
   -\frac{4 \log (z)}{3}
      -\frac{1}{6} z^{49/10} \phi _{J}^2
+\frac{z^5 \left(625 \chi _5-833 \phi _{J} \phi_{V}\right)}{2500}
   -\frac{1}{6} z^{51/10} \phi_{V}^2
   \,+\,{ o}(z^7)
\,.
\eeqs

For $\Delta=13/5$, and hence $\Delta_J=12/5$, $\Delta_V=13/5$, we have
\beqs
\phi&=&
z^{12/5} \phi _{J}+z^{13/5} \phi _{V}
   \,+\,{ o}(z^7)
\,,\\
\chi&=&
   -\frac{\log (z)}{3}
   -\frac{1}{24} z^{24/5} \phi _{J}^2
+\frac{1}{625} z^5 \left(625 \chi _5-52 \phi _{J} \phi_{V}\right)
   -\frac{1}{24} z^{26/5} \phi_{V}^2
   \,+\,{ o}(z^7)
\,,\\
A&=&
   -\frac{4 \log (z)}{3}
   -\frac{1}{6} z^{24/5} \phi _{J}^2
+\frac{z^5 \left(625 \chi _5-832 \phi _{J} \phi_{V}\right)}{2500}
   -\frac{1}{6} z^{26/5} \phi_{V}^2
   \,+\,{ o}(z^7)
\,.
\eeqs

For $\Delta=53/20$, and hence $\Delta_J=47/20$, $\Delta_V=53/20$, we have
\beqs
\phi&=&
z^{47/20} \phi _{J}
+z^{53/20} \phi _{V}
         \,+\,{ o}(z^7)
\,,\\
\chi&=&-\frac{\log (z)}{3}
   -\frac{1}{24} z^{47/10} \phi _{J}^2
   +\frac{z^5 \left(30000 \chi _5-2491 \phi _{J} \phi_{V}\right)}{30000}
   -\frac{1}{24} z^{53/10}  \phi _{V}^2
         \,+\,{ o}(z^7)
\,,\\
A&=&-\frac{4 \log (z)}{3}
   -\frac{1}{6} z^{47/10} \phi _{J}^2
   +\frac{z^5 \left(1875 \chi _5-2491 \phi _{J} \phi_{V}\right)}{7500}
   -\frac{1}{6} z^{53/10}\phi _{V}^2
         \,+\,{ o}(z^7)
\,.
\eeqs

For $\Delta=27/10$, and hence $\Delta_J=23/10$, $\Delta_V=27/10$, we have
\beqs
\phi&=&
z^{23/10} \phi _{J}
+z^{27/10}   \phi _{V}
-\frac{125}{966} z^{69/10} \phi _{J}^3
         \,+\,{ o}(z^7)
\,,\\
\chi&=&-\frac{\log (z)}{3}
   -\frac{1}{24} z^{23/5} \phi _{J}^2
   +\frac{z^5 \left(2500 \chi _5-207 \phi _{J} \phi_{V}\right)}{2500}
   -\frac{1}{24} z^{27/5} \phi _{V}^2
         \,+\,{ o}(z^7)
\,,\\
A&=&-\frac{4 \log (z)}{3}
   -\frac{1}{6} z^{23/5} \phi _{J}^2
   +\frac{z^5 \left(625 \chi _5-828 \phi _{J} \phi_{V}\right)}{2500}
   -\frac{1}{6} z^{27/5} \phi_{V}^2
         \,+\,{ o}(z^7)
\,.
\eeqs

For $\Delta=11/4$, and hence $\Delta_J=9/4$, $\Delta_V=11/4$, we have
\beqs
\phi&=&
z^{9/4} \phi _{J}+z^{11/4} \phi_{V}
-\frac{25}{144} z^{27/4} \phi _{J}^3
         \,+\,{ o}(z^7)
\,,\\
\chi&=&-\frac{\log (z)}{3}
-\frac{1}{24} z^{9/2} \phi _{J}^2
+\frac{1}{400} z^5 \left(400 \chi _5-33 \phi _{J} \phi_{V}\right)
   -\frac{1}{24} z^{11/2} \phi_{V}^2
         \,+\,{ o}(z^7)
\,,\\
A&=&-\frac{4 \log (z)}{3}
-\frac{1}{6} z^{9/2} \phi _{J}^2
+\frac{1}{100} z^5 \left(25 \chi _5-33 \phi _{J} \phi_{V}\right)
   -\frac{1}{6} z^{11/2} \phi_{V}^2
         \,+\,{ o}(z^7)
\,.
\eeqs

{
For $\Delta=3$, and hence $\Delta_J=2$, $\Delta_V=3$, we have
\beqs
    \phi&=& z^2 \phi _J+z^3 \phi _V-\frac{25}{48} z^6 \phi _J^3-\frac{57}{80} z^7
   \left(\phi _J^2 \phi _V\right)+ o(z^7) \,, \\
    \chi&=& -\frac{\log (z)}{3}-\frac{1}{24} z^4 \phi _J^2+z^5 \left(\text{$\chi
   $5}-\frac{2 \phi _J \phi _V}{25}\right)-\frac{1}{24} z^6 \phi
   _V^2+ o(z^7) \,, \\
    A&=& -\frac{4 \log (z)}{3}-\frac{1}{6} z^4 \phi _J^2+z^5
   \left(\frac{\text{$\chi $5}}{4}-\frac{8 \phi _J \phi
   _V}{25}\right)-\frac{1}{6} z^6 \phi _V^2+ o(z^7) \,.
\eeqs
}

For $\Delta=7/2$, and hence $\Delta_J=3/2$, $\Delta_V=7/2$, we have
\beqs
\phi&=&
z^{3/2} \phi _{J}+z^{7/2} \phi _{V}
-\frac{25}{6} z^{9/2} \phi_{J}^3
-\frac{21}{10} z^{13/2} \phi _{J}^2 \phi _{V}
         \,+\,{ o}(z^7)
\,,\\
\chi&=&-\frac{\log (z)}{3}-\frac{1}{24} z^3 \phi_{J}^2
+\frac{1}{100} z^5 \left(100 \chi _5-7 \phi _{J} \phi _{V}\right)
+\frac{25}{96} z^6 \phi _{J}^4
   -\frac{1}{24} z^7 \phi _{V}^2
         \,+\,{ o}(z^7)
\,,\\
A&=&-\frac{4 \log (z)}{3} {-\frac{1}{6} z^3 \phi_{J}^2
+\frac{1}{100} z^5 \left(25 \chi _5-28 \phi _{J} \phi_{V}\right) }
+\frac{25}{24} z^6 \phi _{J}^4
   -\frac{1}{6} z^7 \phi _{V}^2
         \,+\,{ o}(z^7)
\,.
\eeqs

For $\Delta=4$, and hence $\Delta_J=1$, $\Delta_V=4$, we have
\beqs
\phi&=&
z \phi _{J}
+\frac{75}{8} z^3 \phi_{J}^3
+z^4 \phi _{V}
-\frac{7875}{64} z^5 \phi _{J}^5
-\frac{51}{10} z^6 \phi _{J}^2 \phi _{V}
+\frac{183125 z^7 \phi_{J}^7}{1536}
   \,+\,{ o}(z^7)
\,,\\
\chi&=&
-\frac{\log (z)}{3}
-\frac{1}{24} z^2 \phi _{J}^2
-\frac{75}{128} z^4 \phi_{J}^4
+\frac{1}{75} z^5 \left(75 \chi _5-4 \phi _{J} \phi_{V}\right)
+\frac{3125 z^6 \phi _{J}^6}{1536}\\
&&\nonumber
+\frac{1}{280} z^7 \left(125 \chi _5 \phi_{J}^2
-156 \phi _{J}^3 \phi_{V}\right)
   \,+\,{ o}(z^7)
\,,\\
A&=&
-\frac{4 \log (z)}{3}
-\frac{1}{6} z^2 \phi _{J}^2
-\frac{75}{32} z^4 \phi_{J}^4
+\frac{1}{300} z^5 \left(75 \chi _5-64 \phi _{J} \phi_{V}\right)
+\frac{3125}{384} z^6 \phi _{J}^6\\
&&\nonumber
+\frac{z^7 \left(125 \chi _5 \phi _{J}^2-2496 \phi _{J}^3 \phi_{V}\right)}{1120}
   \,+\,{ o}(z^7)
\,.
\eeqs


For $\Delta=24/5$, and hence $\Delta_J=1/5$, $\Delta_V=24/5$, we have
\beqs
\phi&=&
z^{1/5} \phi _{J}+\frac{2875}{168} z^{3/5} \phi _{J} ^3+\frac{589375 z \phi _{J} ^5}{1216}+\frac{1194097540625 z^{7/5} \phi _{J} ^7}{72930816}+\frac{40083780796875 z^{9/5} \phi _{J} ^9}{64827392}\nonumber\\
&&+\frac{206305363012094528125 z^{11/5} \phi _{J} ^{11}}{8070232375296}+\frac{51555188135555851650546875 z^{13/5} \phi _{J} ^{13}}{44741368288641024}\nonumber \\
&&+\frac{220895018700707881287319140625 z^3 \phi _{J} ^{15}}{3872162055526023168}+\frac{143635271586199611133308619908203125 z^{17/5} \phi _{J} ^{17}}{45319784697876575158272}\nonumber\\
&&+\frac{2030227399573934911294696548892509765625 z^{19/5} \phi _{J} ^{19}}{9789073494741340234186752}\nonumber\\
&&+\frac{62734499516892742142454970327066041783203125 z^{21/5} \phi _{J} ^{21}}{3471858066134928669724901376}\nonumber\\
&&+\frac{703737750200907787167436115783648540622242333984375 z^{23/5} \phi _{J} ^{23}}{202562087010576278306429645881344})\nonumber\\
&&+z^{24/5} \phi _{V}-\frac{42889968257354026233434739048798684916940287126220703125 z^5 \phi _{J} ^{25}}{102091291853330444266440541524197376}\nonumber\\
&&-\frac{5313}{125} z^{26/5} \left(\phi _{J} ^2 \phi _{V} \right)+\frac{40372906759830069883219591463928127456136451751634521484375 z^{27/5} \phi _{J} ^{27}}{5811221188951714835775331565278593024}\nonumber\\
&&+\frac{512647 z^{28/5} \phi _{J} ^4 \phi _{V} }{3024}+\frac{23451902891374040266318476466845032509323071584144748101806640625 z^{29/5} \phi _{J} ^{29}}{1344344664967333903248922102985288595800064}\nonumber\\
&& +\frac{2812375255 z^6 \phi _{J} ^6 \phi _{V} }{6664896}-\frac{9356448978981275 z^{32/5} \left(\phi _{J} ^8 \phi _{V} \right)}{786777643008}\nonumber\\
&&+\frac{3382994549224018067073083754105413390854841531398992869892655029296875 z^{31/5} \phi _{J} ^{31}}{12873444511727189457511678058187123593381412864}\nonumber\\
&&+\frac{63252672595173277436174439987322783944040397901181003729679595947265625 z^{33/5} \phi _{J} ^{33}}{11443061788201946184454824940610776527450144768}\nonumber\\
&&-\frac{882653673174311913125 z^{34/5} \left(\phi _{J} ^{10} \phi _{V} \right)}{1090473813209088}
         \,+\,{ \mathcal O(z^7)}
\,,\\
\chi&=&-\frac{\log (z)}{3}-\frac{1}{24} \phi _{J} ^2 z^{2/5}-\frac{2875 \phi _{J} ^4
   z^{4/5}}{2688}-\frac{445840625 \phi _{J} ^6
   z^{6/5}}{12870144}-\frac{1245177090625 \phi _{J} ^8
   z^{8/5}}{1000194048}\nonumber\\
&&-\frac{12757500935258125 \phi _{J} ^{10}
   z^2}{266051616768}-\frac{252462758077427834375 \phi _{J} ^{12}
   z^{12/5}}{129123718004736}\nonumber\\
&&-\frac{219400845248171582865859375 \phi _{J} ^{14}
   z^{14/5}}{2607782608823648256}-\frac{595634988671267098623623224609375 \phi_J ^{16}
   z^{16/5}}{155382118964148257685504}\nonumber\\
&&-\frac{37370923454539023891704902220703125
   \phi _{J} ^{18}
   z^{18/5}}{199777010096762045595648}-\frac{30055898516617546756330722827491167968
   75 \phi _{J} ^{20}
   z^4}{299011699475735483516977152}\nonumber\\
&&-\frac{3906039992092721696123052068906280373737
   55859375 \phi _{J} ^{22}
   z^{22/5}}{607686261031728834919288937644032}\nonumber\\
&&-\frac{13963985971796191572802079948
   45812226729262841796875 \phi _{J} ^{24}
   z^{24/5}}{19445960353015322717417246004609024}\nonumber\\
&&+\left(-\frac{1}{625} (8 \phi_{J}  \phi _{V} )+\chi_5\right)
   z^5+\frac{1306083301937461832252797063598660554434038579290771484375 \phi _{J}
   ^{26} z^{26/5}}{605197178106542873611459530155442044928}\nonumber\\
&&+\left(-\frac{4117 \phi_{J} ^3 \phi _{V} }{70875}+\frac{125 \phi _{J} ^2 \chi_5}{216}\right)
   z^{27/5}+\frac{5
   \phi _{J} ^4 (-9784936 \phi _{J}  \phi _{V} +55981125 \chi_5)
   z^{29/5}}{19994688}\nonumber\\
&&+\frac{13030865310005017933330153469549477844134773156086490478515625
   \phi _{J} ^{28} z^{28/5}}{474474587635529612911384271641866563223552}\nonumber\\
&&+\frac{703276966570738727880625530782313795347013255222415921
   8505859375 \phi _{J} ^{30}
   z^6}{10754757319738671225991376823882308766400512}\nonumber\\
&&-\frac{25 \left(\phi _{J} ^6
   (2692663154104 \phi _{J}  \phi _{V} -10082616060375 \chi_5)\right)
   z^{31/5}}{590083232256}\nonumber\\
&&+\frac{55409694774380233135646018320740443010540036633898
   67568262069091796875 \phi _{J} ^{32}
   z^{32/5}}{290788393676661220687322610255520909403438972928}\nonumber\\
&&-\frac{25 \left(\phi_{J} ^8 (35613091214958616 \phi _{J}  \phi _{V} -90309329147491875
   \chi_5)\right)
   z^{33/5}}{155781973315584}\nonumber\\
&&+\frac{18183027897024617096562220920784271134126559768
   5817924668243991119384765625 \phi _{J} ^{34}
   z^{34/5}}{294132460203942824725226820273459399861578521116672}\nonumber\\
&&
         \,+\,{ \mathcal O(z^7)}
\,,\\
A&=&-\frac{4 \log (z)}{3}-\frac{1}{6} \phi _{J} ^2 z^{2/5}-\frac{2875}{672} \phi _{J} ^4
   z^{4/5}-\frac{445840625 \phi _{J} ^6 z^{6/5}}{3217536}-\frac{1245177090625
   \phi _{J} ^8 z^{8/5}}{250048512}\nonumber\\
&&-\frac{12757500935258125 \phi _{J} ^{10}
   z^2}{66512904192}-\frac{252462758077427834375 \phi _{J} ^{12}
   z^{12/5}}{32280929501184}\nonumber\\
&&-\frac{219400845248171582865859375 \phi _{J} ^{14}
   z^{14/5}}{651945652205912064}-\frac{595634988671267098623623224609375 \phi_{J} ^{16}
   z^{16/5}}{38845529741037064421376}\nonumber\\
&&-\frac{37370923454539023891704902220703125
   \phi _{J} ^{18}
   z^{18/5}}{49944252524190511398912}-\frac{300558985166175467563307228274911679687
   5 \phi _{J} ^{20}
   z^4}{74752924868933870879244288}\nonumber\\
&&-\frac{39060399920927216961230520689062803737375
   5859375 \phi _{J} ^{22}
   z^{22/5}}{151921565257932208729822234411008}\nonumber\\
&&-\frac{13963985971796191572802079948
   45812226729262841796875 \phi _{J} ^{24}
   z^{24/5}}{4861490088253830679354311501152256}\nonumber\\
&&+\left(-\frac{1}{625} (32 \phi_{J}  \phi _{V} )+\frac{\chi_5}{4}\right)
   z^5+\frac{1306083301937461832252797063598660554434038579290771484375 \phi _{J}
   ^{26} z^{26/5}}{151299294526635718402864882538860511232}\nonumber\\
&&+\left(-\frac{16468 \phi_{J} ^3 \phi _{V} }{70875}+\frac{125 \phi _{J} ^2 \chi_5}{864}\right)
   z^{27/5}+\left(-\frac{873655 \phi_{J} ^5 \phi _{V} }{89262}+\frac{181875 \phi _{J} ^4 \chi_5}{51968}\right)
   z^{29/5}\nonumber\\
&&+\frac{13030865310005017933330153469549477844134773156086490478515625
   \phi _{J} ^{28}
   z^{28/5}}{118618646908882403227846067910466640805888}\nonumber\\
&&+\frac{7032769665707387278806255307823137953470132552224159218505859375
   \phi _{J} ^{30} z^6}{2688689329934667806497844205970577191600128}\nonumber\\
&&+\frac{25
   \phi _{J} ^6 (-43082610465664 \phi _{J}  \phi _{V} +10082616060375
   \chi_5)
   z^{31/5}}{2360332929024}\nonumber\\
&&+\frac{5540969477438023313564601832074044301054003663389
   867568262069091796875 \phi _{J} ^{32}
   z^{32/5}}{72697098419165305171830652563880227350859743232}\nonumber\\
&&-\frac{25 \left(\phi_{J} ^8 (569809459439337856 \phi _{J}  \phi _{V} -90309329147491875
   \chi_5)\right)
   z^{33/5}}{623127893262336}\nonumber\\
&&+\frac{18183027897024617096562220920784271134126559768
   5817924668243991119384765625 \phi _{J} ^{34}
   z^{34/5}}{73533115050985706181306705068364849965394630279168}\nonumber\\
&&
         \,+\, { \mathcal O(z^7)}
\,.
\eeqs


\section{Asymptotic expansions of the fluctuations}
\label{sec:IRUVexpansions}

As described in the body of the paper, in order to improve the numerical convergence of the 
spectra of fluctuations, we write the asymptotic expansions of such solutions,
both in the IR and in the UV.
For the UV case, the expansion depends crucially on $\Delta$; while we did compute it explicitly for
all the values of $\Delta$ used in the numerical study, we report here only one example, 
for illustrative purposes.

\subsection{IR expansions}

For convenience, we put $\rho_o = 0$ and $A_I = 0$ in this subsection,\footnote{The dependence on $\rho_o$ and $A_I$ can be reinstated by making the substitutions $\rho \rightarrow \rho - \rho_o$ and $q^2 \rightarrow e^{-2A_I} q^2$ in the expressions.} while $\chi_I = 0$ in order to avoid a conical singularity, as explained in the main text.
For the fluctuations of the scalars $\Phi^a=\{ {\phi},\chi \}$, we have
\begin{align}
\mathfrak a^1 =&\ \mathfrak a^1_{I,0}+\mathfrak a^1_{I,l} \log (\rho )+\frac{1}{4} \rho ^2 \bigg[ -\frac{1}{4} \Delta  \left(\mathfrak a^1_{I,0} \left(\Delta  \left(15 \phi_I^2-4\right)+20\right)+6 \phi_I (\mathfrak a^2_{I,0}-\mathfrak a^2_{I,l}) \left(\Delta  \left(5 \phi_I^2-4\right)+20\right)\right)
\nonumber \\ &
+q^2 (\mathfrak a^1_{I,0}-\mathfrak a^1_{I,l}) -\frac{1}{48} \mathfrak a^1_{I,l} \left(\Delta  \left(25 \Delta  \phi_I^4+20 (10-11 \Delta ) \phi_I^2+48 (\Delta -5)\right)+400\right)
\nonumber \\ &
+\log (\rho ) \left(\mathfrak a^1_{I,l} \left(-\frac{15 \Delta ^2 \phi_I^2}{4}+(\Delta -5) \Delta +q^2\right)-\frac{3}{2} \mathfrak a^2_{I,l} \Delta  \phi_I \left(\Delta  \left(5 \phi_I^2-4\right)+20\right)\right) \bigg]+\mathcal O \left(\rho ^4\right) \,, \\ 
\mathfrak a^2 =&\ \mathfrak a^2_{I,0}+\mathfrak a^2_{I,l} \log (\rho )+\frac{1}{4} \rho ^2 \bigg[-\frac{1}{4} \Delta  \phi_I (\mathfrak a^1_{I,0}-\mathfrak a^1_{I,l}) \left(\Delta  \left(5 \phi_I^2-4\right)+20\right) +q^2 (\mathfrak a^2_{I,0}-\mathfrak a^2_{I,l})
\nonumber \\ &
-\frac{3}{8} \mathfrak a^2_{I,0} \left(\Delta  \phi_I^2 \left(\Delta  \left(5 \phi_I^2-8\right)+40\right)+80\right)+\frac{13}{48} \mathfrak a^2_{I,l} \left(\Delta  \phi_I^2 \left(\Delta  \left(5 \phi_I^2-8\right)+40\right)+80\right)
\nonumber \\ &
+\log (\rho ) \left(-\frac{5}{4} \mathfrak a^1_{I,l} \Delta ^2 \phi_I^3+\mathfrak a^1_{I,l} (\Delta -5) \Delta  \phi_I+\mathfrak a^2_{I,l} \left(-\frac{15}{8} \Delta ^2 \phi_I^4+3 (\Delta -5) \Delta  \phi_I^2+q^2-30\right)\right)\bigg]+\mathcal O \left(\rho ^4\right) \,.
\end{align}

For the fluctuations of the vector $\chi_M$, we have
\begin{align}
\mathfrak v =&\ \mathfrak v_{I,-2} \rho ^{-2}+\frac{1}{2} q^2 \mathfrak v_{I,-2} \log (\rho )+\mathfrak v_{I,0} +\frac{1}{12288} \rho ^2 \Big[1536 q^2 \mathfrak v_{I,0}+80 \Delta ^2 \mathfrak v_{I,-2} \phi_I^4 \left(2 \left(8 \Delta ^2-50 \Delta +75\right)-3 q^2\right)
\nonumber \\ &
+128 (\Delta -5) \Delta  \mathfrak v_{I,-2} \phi_I^2 \left(-3 (\Delta -5) \Delta +3 q^2-50\right)-64 \left(9 q^4+60 q^2-500\right) \mathfrak v_{I,-2}+125 \Delta ^4 \mathfrak v_{I,-2} \phi_I^8
\nonumber \\ &
-1000 (\Delta -2) \Delta ^3 \mathfrak v_{I,-2} \phi_I^6+ 768 q^4 \mathfrak v_{I,-2} \log (\rho )\Big]+\mathcal O \left(\rho ^4\right)  \,.
\end{align}

For the tensor fluctuations $\mathfrak e^{\mu}_{\ \nu}$, we have
\beq
\mathfrak e = \mathfrak e_{I,0}+\mathfrak e_{I,l} \log (\rho )+\frac{1}{192} \rho ^2 \Big[48 q^2 (\mathfrak e_{I,0}-\mathfrak e_{I,l})-25 \Delta ^2 \mathfrak e_{I,l} \phi_I^4+40 (\Delta -5) \Delta  \mathfrak e_{I,l} \phi_I^2-400 \mathfrak e_{I,l}+48 \mathfrak e_{I,l} q^2 \log (\rho )\Big]+\mathcal O \left(\rho ^4\right) \,.
\eeq

\subsection{UV expansions}

In this subsection, we put $\Delta = 3$, and $A_U = 0 = \chi_U$.\footnote{The dependence on $\chi_U$ and $A_U$ can be reinstated by making the substitution $q^2 \rightarrow e^{2\chi_U-2A_U} q^2$ in the expressions.} We write the expansions in terms of $z \equiv e^{-\rho}$.

For the fluctuations of the scalars, we have
\begin{align}
\mathfrak a^1 =&\ \mathfrak a^1_2 z^2+\mathfrak a^1_3 z^3+\frac{1}{2} \mathfrak a^1_2 q^2 z^4+\frac{1}{6} \mathfrak a^1_3 q^2 z^5+\frac{1}{48} \mathfrak a^1_2 \left(2 q^4-99 \phi_J^2\right) z^6 +\mathcal O \left(z^7\right) \,, \\ 
\mathfrak a^2 =&\ \mathfrak a^2_0-\frac{1}{6} \mathfrak a^2_0 q^2 z^2 +\frac{1}{24} \mathfrak a^2_0 q^4 z^4+\mathfrak a^2_5 z^5+\frac{1}{144} \mathfrak a^2_0 q^2 \left(q^4-14 \phi_J^2\right) z^6 +\mathcal O \left(z^7\right)\,.
\end{align}

For the vector fluctuations, we have
\begin{align}
\mathfrak v =&\ \mathfrak v_0-\frac{1}{6} q^2 \mathfrak v_0 z^2 +\frac{1}{24} q^4 \mathfrak v_0 z^4+\mathfrak v_5 z^5+\frac{1}{144} q^2 \mathfrak v_0 \left(q^4-14 \phi_J^2\right) z^6
\nonumber \\ &
+\frac{1}{70} q^2 (70 \mathfrak v_0 \chi_5-2 \mathfrak v_0 \phi_J \phi_V+5 \mathfrak v_5) z^7 +\mathcal O \left(z^8\right) \,.
\end{align}

For the tensor fluctuations, we have
\beq
\mathfrak e = \mathfrak e_0- \frac{1}{6} \mathfrak e_0 q^2 z^2 +\frac{1}{24} \mathfrak e_0 q^4 z^4+\mathfrak e_5 z^5+\mathcal O \left(z^6\right) \,.
\eeq

\section{Probe approximation}
\label{sec:probedetail}
In this appendix, we further discuss the probe approximation. More details can be found in Ref.~\cite{Elander:2020csd}. The fluctuations $\mathfrak{a}^a=\varphi^a\,-\,\frac{\partial_r\Phi^a}{6\partial_r A} h$ are formed by the mixing of the fluctuations of the scalars $\Phi^a=({\phi},\chi)$ and fluctuations $h$ of the trace of the four-dimensional part of the metric. The latter couples to the trace of the stress-energy tensor of the boundary theory. If $\mathfrak{a}^a$ is mostly composed of $h$, hence  $\mathfrak{a}^a \simeq \frac{\partial_r\Phi^a}{6\partial_r A} h$, the couplings of the states sourced by the operators at the boundary are well approximated by the dilatonic counterparts. 
Conversely, in the probe approximation for computing spectra, one neglects the mixing between the scalar and metric fluctuations.
This will lead to correct results only if $\mathfrak{a}^a \simeq \varphi^a$, such that the contribution of $h$ in Eq.~(\ref{eq:Ascalarflucs1}) can be neglected.

The probe approximation is hence obtained by neglecting the contributions from the metric fluctuations, in particular $h$, in Eqs.~\eqref{eq:Ascalarflucs1} and~\eqref{eq:Ascalarflucs3}, leading to~\cite{Elander:2020csd}
\beqs
0&=&\left[\frac{}{}{\cal D}_r^2 + (D-1)\partial_r A\,{\cal D}_r -e^{-2A} q^2\right]\mathfrak{p}^a-\left[\frac{}{}V^a_{\,\,\,\,|c}-{\cal R}^a_{\,\,\,\,bcd}\partial_r\Phi^b\partial_r \Phi^d
\right]\mathfrak{p}^c\,,
\label{eq:probe}
\eeqs
for the equations of motion, while the boundary conditions reduce to
\beqs
0&=&\left.\frac{}{}\mathfrak{p}^a\right|_{r_i}\,.
\label{eq:probeboundary}
\eeqs
In these equations, we have replaced $\mathfrak a^a$ with the probe fluctuations that we denote by $\mathfrak p^a$.

We perform the calculation of the spectra of scalar fluctuations in two ways for $\Delta=5/2$ in the body of the paper.
First, by solving the exact Eqs.~(\ref{eq:Ascalarflucs1}), with the boundary 
conditions in Eqs.~(\ref{eq:Ascalarflucs3}), and finding the spectrum of masses.
Then, we repeat the calculation for the same background, but using the probe approximation
and solving Eqs.~(\ref{eq:probe}), with the boundary conditions in Eqs.~(\ref{eq:probeboundary}).
If the two processes result in different spectra, overlap with the dilaton can not be neglected. 
In the next subsections, IR and UV expansions for fluctuations in the probe approximation are given.

\subsection{IR expansions}
In this subsection, we put $\Delta = 5/2$. As in Appendix~\ref{sec:IRUVexpansions}, we put $\rho_o = 0$ and $A_I = 0$, while $\chi_I = 0$ in order to avoid a conical singularity.
For the fluctuations of the scalars in the probe approximation, we have
\begin{align}
\mathfrak p^1 = \,& \mathfrak p^1_{I,0}+\mathfrak p^1_{I,l} \log (\rho )+\rho ^2 \bigg[ \Delta  \left(-\frac{5 \mathfrak p^2_{I,l} \phi_I}{2}+\frac{5 \mathfrak p^2_{I,0} \phi_I}{2}\right)+ \Delta ^2
   \left(\frac{\mathfrak p^2_{I,l} \phi_I}{2}-\frac{\mathfrak p^2_{I,0} \phi_I}{2}-\frac{5 \mathfrak p^2_{I,l} \phi_I^3}{8}+\frac{5 \mathfrak p^2_{I,0} \phi_I^3}{8}\right)+\nonumber \\ 
   & \mathfrak p^1_{I,0} \left(\frac{q^2}{4}-\frac{5 \Delta }{4}+\Delta ^2 \left(\frac{1}{4}-\frac{15 \phi_I^2}{16}\right)\right)+ 
   \mathfrak p^1_{I,l}\left(-\frac{25}{12}-\frac{q^2}{4}+\Delta  \left(\frac{5}{4}-\frac{25 \phi_I^2}{24}\right)+\Delta ^2 \left(-\frac{1}{4}+\frac{55 \phi_I^2}{48}-\frac{25 \phi_I^4}{192}\right)\right)+ \nonumber \\ 
   &\left(\frac{5 \mathfrak p^2_{I,l} \Delta  \phi_I}{2}+\Delta ^2 \left(-\frac{\mathfrak p^2_{I,l} \phi_I}{2} +\frac{5 \mathfrak p^2_{I,l} \phi_I^3}{8}\right)+\mathfrak p^1_{I,l} \left(\frac{q^2}{4}-\frac{5 \Delta }{4}+\Delta ^2 \left(\frac{1}{4}-\frac{15 \phi_I^2}{16}\right)\right)\right) \log (\rho )\bigg]+\mathcal O \left(\rho ^4\right) \,, \\
   \mathfrak p^2 = \,& \mathfrak p^2_{I,0}+\mathfrak p^2_{I,l} \log (\rho )+\rho ^2 \bigg[\Delta  \left(-\frac{5 \mathfrak p^1_{I,l} \phi_I}{12} +\frac{5 \mathfrak p^1_{I,0} \phi_I}{12}\right)+\Delta ^2
   \left(\frac{\mathfrak p^1_{I,l} \phi_I}{12}-\frac{\mathfrak p^1_{I,0} \phi_I}{12}-\frac{5 \mathfrak p^1_{I,l} \phi_I^3}{48}+\frac{5 \mathfrak p^1_{I,0} \phi_I^3}{48}\right)+\nonumber \\ 
   &\mathfrak p^2_{I,l} \left(-\frac{5}{4}-\frac{q^2}{4}-\frac{5 \Delta  \phi_I^2}{8}+\Delta ^2 \left(\frac{\phi_I^2}{8}-\frac{5 \phi_I^4}{64}\right)\right)+\mathfrak p^2_{I,0} \left(-\frac{5}{6}+\frac{q^2}{4}-\frac{5 \Delta  \phi_I^2}{12}+\Delta ^2 \left(\frac{\phi_I^2}{12}-\frac{5\phi_I^4}{96}\right)\right)+\nonumber \\ 
   & \hspace{-1cm} \left(\frac{5 \mathfrak p^1_{I,l} \Delta  \phi_I}{12}+\Delta ^2 \left(-\frac{\mathfrak p^1_{I,l} \phi_I}{12} +\frac{5
   \mathfrak p^1_{I,l} \phi_I^3}{48}\right)+\mathfrak p^2_{I,l} \left(-\frac{5}{6}+\frac{q^2}{4}-\frac{5 \Delta  \phi_I^2}{12}+\Delta ^2 \left(\frac{\phi_I^2}{12}-\frac{5 \phi_I^4}{96}\right)\right)\right) \log (\rho )\bigg]+\mathcal O \left(\rho ^4\right) \,.
\end{align}

\subsection{UV expansions}

For the UV expansions we put $A_U = 0 = \chi_U$, as explained in Appendix~\ref{sec:IRUVexpansions}. Also, the expansions are considered for $\Delta = 5/2$ and are written in terms of $z \equiv e^{-\rho}$.

The fluctuations of the scalars in the probe approximation are more complicated to derive in this case as some of the exponents in expansions are non-rational numbers. 
The part of the expansion which is of importance in our calculations will be in the following form:
\beqs
\mathfrak p^1 &=& \log(z) z^{5/2} \mathfrak p^1_{5/2,l} (1 + \cdots) + z^{5/2} \mathfrak p^1_{5/2} (1 + \cdots) \,, \\ 
\mathfrak p^2 &=& z^{\alpha_-} \mathfrak p^2_- (1 + \cdots) + z^{\alpha_+} \mathfrak p^2_+ (1 + \cdots) \,,
\eeqs
where $\alpha_\pm \equiv \big( 5 \pm \sqrt{35/3} \big)/2$.

\section{More about the free energy}
\label{sec:Fdetail}

In this short Appendix, we report some details of the  free energy density study for values of $\Delta$ close to $\Delta \simeq 1.8$.
In Fig.~\ref{Fig:FvsJ_4zoom} we show a detail of Fig.~\ref{Fig:FvsJ_4}, that highlights the evolution of 
the free energy of the confining solutions. We notice that this is monotonic for $\Delta=1.7$,
while it has a local maximum and minimum, and changes in concavity, when $\Delta = 1.95$.
It would be tempting to correlate this behaviour with the appearance of a tachyon in the spectrum,
but, as we comment in the main body of the paper, the presence of divergences, treated via
holographic renormalisation, turns the concavity of ${\cal F}$ into a scheme-dependent quantity.
We hence do not further pursue this line of enquiry in this paper.

\begin{figure}[th]
\begin{center}
\includegraphics[width=17cm]{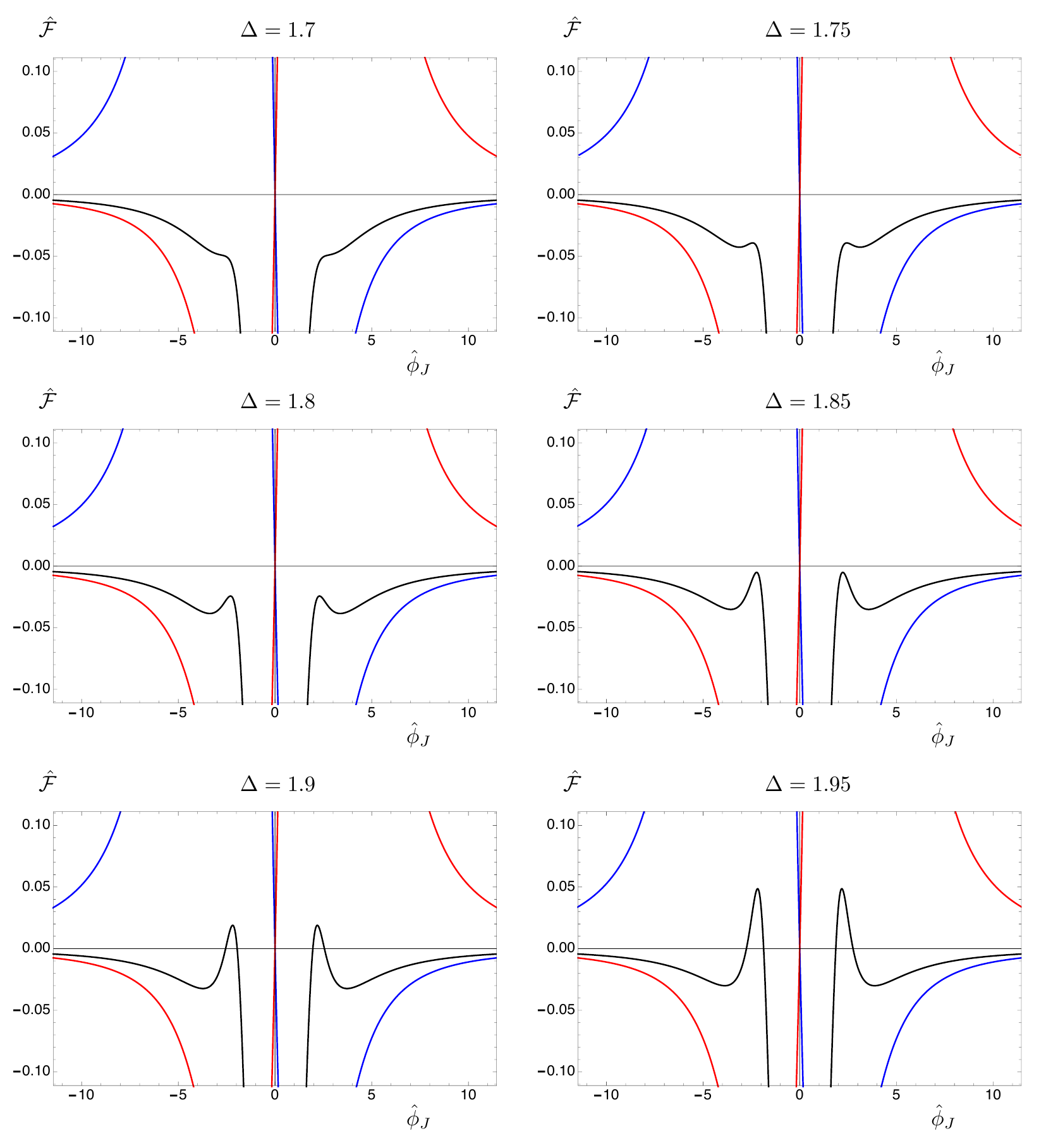}
\caption{The free energy $\hat{\cal F}$ as a function of the source $\hat{\phi}_J$, in units of the scale $\Lambda$, for representative choices of $\Delta$
near $\Delta\simeq 1.8$, above which the spectrum of fluctuations contains a tachyon in part of the parameter space. The black curve denotes the confining solutions, while the red and blue solutions are singular 
domain-wall ones.}
\label{Fig:FvsJ_4zoom}
\end{center}
\end{figure}

\newpage

\end{document}